\documentstyle[aps,pre,epsf]{revtex}

\newcommand{\bfr}{{\bf r}} % keep all these braces...

\newcommand{\arcsinh}{\mathop{\rm arcsinh}\nolimits}
\newcommand{\arctanh}{\mathop{\rm arctanh}\nolimits}
\renewcommand{\epsilon}{\varepsilon}

\begin{document}
\draft % makes pacs numbers print
\title{Crossover scaling in two dimensions}
\author{Erik Luijten\thanks{Electronic address: erik@tntnhb3.tn.tudelft.nl}
        and Henk~W.~J. Bl\"ote}
\address{Department of Physics, Delft University of Technology,
         Lorentzweg 1, 2628 CJ Delft, The Netherlands}
\author{Kurt Binder}
\address{Institut f\"ur Physik, WA 331, Johannes Gutenberg-Universit\"at,
  D-55099 Mainz, Germany}

\date{\today}
\maketitle
\begin{abstract}
  We determine the scaling functions describing the crossover from Ising-like
  critical behavior to classical critical behavior in two-dimensional systems
  with a variable interaction range. Since this crossover spans several decades
  in the reduced temperature as well as in the finite-size crossover variable,
  it has up to now largely evaded a satisfactory numerical determination. Using
  a new Monte Carlo method, we could obtain accurate results for sufficiently
  large interactions ranges.  Our data cover the full crossover region both
  above and below the critical temperature and support the hypothesis that the
  crossover functions are universal. Also the so-called effective exponents are
  discussed and we show that these can vary nonmonotonically in the crossover
  region.
\end{abstract}

%\narrowtext

\section{Introduction}
The crossover from Ising-like to classical critical behavior has attracted
renewed attention in recent years. This crossover behavior occurs in many
thermodynamic systems, such as ionic solutions, simple fluids, fluid mixtures,
and polymer mixtures. The Ginzburg criterion~\cite{ginzburg} states that
sufficiently close to the critical point these systems exhibit critical
exponents belonging to the 3D Ising universality class. At larger distances
from the critical point, but still within the critical region, classical
(mean-field-like) critical exponents are observed. Although this appears to be
a well-established picture, the precise nature of the crossover between these
two universality classes is still subject to investigation.  For example,
Anisimov {\em et al.\/} recently claimed~\cite{anisimov95} to have observed an
``effective'' susceptibility exponent which varied {\em nonmonotonically\/}
from its classical value $\gamma_{\rm MF}=1$ to its Ising value $\gamma_{\rm I}
\approx 1.24$ when the critical point was approached.  Later, the possibility
of such behavior within the critical domain was questioned by Bagnuls and
Bervillier, see Refs.~\cite{bagnuls,anisimov96}.  On the other hand, Fisher has
argued~\cite{fisher-eff} that nonmonotonical variation of effective critical
exponents is not necessarily an indication of nonuniversal behavior.  Other
questions concern the size of the crossover region, which is expected to span
several decades in the crossover variable~\cite{anisimov92}, and the size of
the temperature region around $T_{\rm c}$ within which Ising-like behavior is
observed~\cite{fisherlee}.  Until now it has turned out to be very difficult to
accurately observe the full crossover region in numerical simulations. A major
effort has been undertaken in Ref.~\cite{deutsch93} for three-dimensional
polymer mixtures, where crossover occurs as a function of the polymer chain
length. However, despite chain lengths of up to 512~monomers, the results did
not span the full crossover region. For this reason, Mon and Binder~\cite{mon}
turned their attention to the two-dimensional Ising model with an extended
range of interaction, where a crossover from Ising-like to classical critical
behavior occurs when the range $R$ of the spin--spin interactions is increased
(suppressing the critical fluctuations).  In two dimensions one can not only
access larger interaction ranges, but also both asymptotic regimes are known
exactly and the variation of the critical exponents is considerably larger than
in the crossover from 3D Ising-like critical behavior to classical critical
behavior.  Mon and Binder derived the (singular) $R$~dependence of the critical
amplitudes of scaling functions and carried out Monte Carlo simulations to
verify these predictions numerically.  Even in these two-dimensional systems,
the mean-field regime turned out to be only barely reachable.

In a recent paper~\cite{medran}, we rederived the predictions of Mon and Binder
from renormalization theory and also obtained the $R$~dependence of various
corrections to scaling, such as the shift of the critical temperature with
respect to the mean-field critical temperature.  Furthermore, larger
interaction ranges and system sizes were accessible to our numerical
simulations thanks to a dedicated Monte Carlo algorithm. This enabled us to
actually verify the theoretical predictions in two-dimensional systems. In this
paper, we show that the simulations presented in Ref.~\cite{medran} allow a
full mapping of the finite-size crossover curves for various quantities.
However, these curves describe the {\em finite-size\/} dependences of critical
amplitudes, which have (to our knowledge) not been observed experimentally.
Therefore we have also carried out simulations at temperatures further from the
critical temperature in order to observe the {\em thermal\/} crossover of these
quantities. The results of these simulations, which partially have been
reported in an earlier note~\cite{chicross}, are presented as well.  The fact
that in our model both the temperature distance from the critical point and the
interaction range can be varied turns out to be essential to observe the full
crossover region.

The outline of the remainder this paper is as follows. After a short
recapitulation of the model under investigation (Sec.~\ref{sec:model}) we start
in Sec.~\ref{sec:finite} with finite-size crossover scaling.  We discuss the
required system sizes and interaction ranges and obtain the crossover curves
for the absolute magnetization density, magnetic susceptibility, and the
spin--spin correlation function over half the system size. Thermal crossover
scaling is treated in Sec.~\ref{sec:thermal}, where we consider the approach of
$T_{\rm c}$ both in the symmetric phase ($T>T_{\rm c}$) and in the state of
broken symmetry ($T<T_{\rm c}$).  Again, crossover curves are obtained for the
order parameter and the susceptibility. The various aspects of these curves are
discussed in some detail.  Graphs of the logarithmic derivatives of the
crossover curves, which can be associated with so-called effective critical
exponents as measured in experiments, are presented in Sec.~\ref{sec:effexp}.
In Sec.~\ref{sec:conclusions} we end with a summary of our conclusions.

\section{Short description of the model}
\label{sec:model}
Let us first briefly recall the model as it was introduced in Ref.~\cite{mon}.
This is a two-dimensional Ising system consisting of $L \times L$ lattice
sites with periodic boundary conditions. Each spin in the system interacts
equally with its $z$~neighbors lying within a distance $R_m$. This defines
the coupling between two spins $s_i$ and~$s_j$ at a distance~$r$ as
\begin{equation}
K_{ij} = K(r) \equiv \left\{ \begin{array}{ll}
                                cR_m^{-d} & \mbox{if $r \leq R_m$} \\
                                        0 & \mbox{if $r > R_m$.}
                             \end{array}
                     \right.
\end{equation}
In the absence of an external magnetic field the Hamiltonian is
\begin{equation}
{\cal H}/k_{\rm B}T =  -\sum_{i}\sum_{j>i} K(|\bfr_i-\bfr_j|) s_i s_j \;,
\end{equation}
where the sums run over all spins in the system and $\bfr_i$ denotes the
position of spin $s_i$. To suppress lattice effects we use an {\em effective\/}
interaction range~$R$, defined as
\begin{eqnarray}
R^2 &\equiv& \frac{\sum_{j \neq i} (\bfr_i - \bfr_j)^2 K_{ij}}%
                  {\sum_{j \neq i} K_{ij}} \nonumber \\
    &=& \frac{1}{z} \sum_{j \neq i} |\bfr_i - \bfr_j |^2
        \quad\quad \text{with } |\bfr_i - \bfr_j | \leq R_m \;.
\end{eqnarray}
For large ranges, $R$ approaches the limiting value $R_m/\sqrt{2}$.

\section{Finite-size crossover scaling}
\label{sec:finite}

\subsection{General considerations}
It has been shown by Binder and Deutsch~\cite{binder92} that crossover scaling
can be combined with finite-size scaling by including the dependence on the
crossover variable in the probability distribution function of the order
parameter. Indeed, just as crossover in the thermodynamic limit is described as
a function of the reduced temperature divided by the Ginzburg number, it can be
described as the function of a size-dependent crossover variable~$G$ in finite
systems. In Ref.~\cite{mon}, this crossover variable was derived as
$G=LR^{-4/(4-d)}$, where $L$ is the linear system size and $d$ denotes the
dimensionality. This also follows from the renormalization treatment in
Ref.~\cite{medran}. In short, sufficiently close to the Gaussian fixed point
(i.e., for a sufficiently large interaction range $R$) the critical behavior
will be classical. In terms of a renormalized Landau--Ginzburg--Wilson~(LGW)
Hamiltonian in momentum space, this implies that the coefficient of the
$\phi^4$~term must be much smaller than that of the $\phi^2$~term,
$uL^{4-d}/R^{4} \ll 1$ (cf.\ in particular Eq.~(6) of Ref.~\cite{medran}),
which again leads to the crossover parameter $LR^{-4/(4-d)}$, where for the
moment we assume that $u$ is of order unity.

\begin{figure}
\begin{center}
\leavevmode
\epsfbox{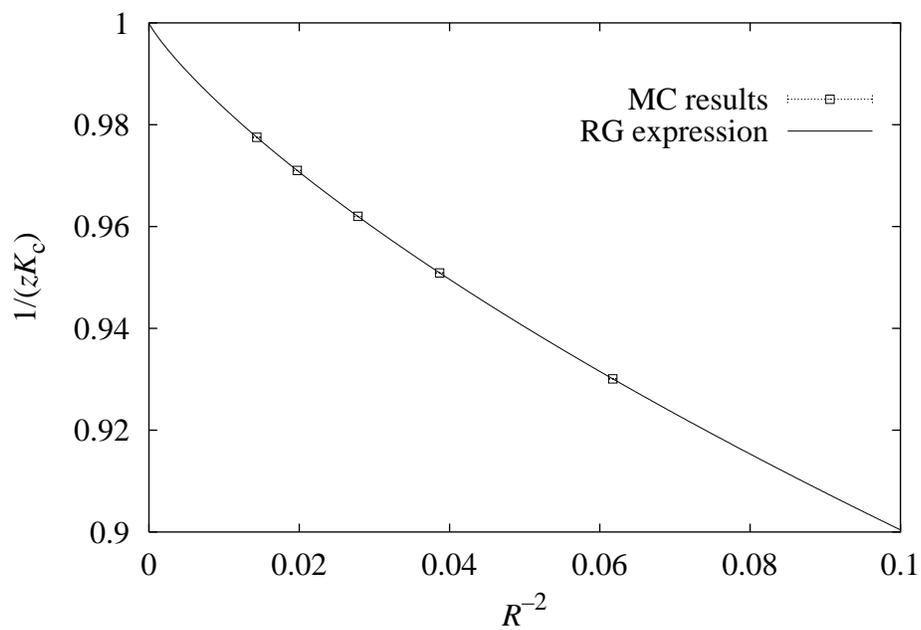}
\end{center}
\caption[]{The critical temperature as a function of the inverse interaction
  range, together with the renormalization
  expression~(\protect\ref{eq:tc-scale}) fitted to it.}
\label{fig:kc-scale}
\end{figure}

In Ref.~\cite{medran}, we focused our attention on the critical finite-size
amplitudes in the limit of $L \to \infty$.  Here we will examine the crossover
in the corresponding data for {\em finite\/} system sizes. Since the crossover
regime is expected~\cite{fisher-eff,anisimov92} to span several decades in the
crossover variable $G=L/R^2$, it is numerically not feasible to observe both
asymptotic regimes by merely varying the system size $L$ while keeping the
range $R$ fixed. Therefore we construct the curve by combining the results for
various values of $R$, cf.\ Ref.~\cite{deutsch93}. Indeed, the Ising regime
($L/R^2 \gg 1$) is easily reachable, although the results for very small ranges
do not conform well to the leading $R$~dependence of the critical scaling
functions~\cite{mon} and are thus, at first sight, not well suited for
constructing the crossover curve.  The mean-field regime ($L/R^2 \ll 1$),
however, poses more substantial problems. If the linear system size~$L$ is made
too small, the numerical results exhibit strong finite-size effects. Therefore
$L$ must be at least of the order of the interaction range.  More precisely,
boundary effects will occur for systems for which $L \approx R_m$ and the
smallest possible value of the crossover variable~$G$ is roughly equal to
$R_m/R^2 \approx \sqrt{2}/R$.  Thus, large ranges are required to reach the
regime where $G \ll 1$.  In a conventional Monte Carlo algorithm, the
efficiency of simulations rapidly decreases with increasing interaction range.
This limitation has been circumvented by applying a dedicated cluster
algorithm, as explained in Ref.~\cite{medran}.  Still, a problem remains.
Namely, the finite-size crossover scaling is valid {\em at\/} the critical
temperature. Any deviation from this temperature will lead to systematic errors
in the analysis. Since the (range-dependent) critical temperatures are
determined in the Ising limit, i.e.\ from system sizes $L > R^2$, large
interaction ranges require {\em very\/} large system sizes for an accurate
determination of $T_{\rm c}$. For example, the most efficient way to obtain
data for $G \approx 0.02$ is to simulate a system with $L=100$ and $R_m=100$
($R \approx 70$). However, an accurate determination of $T_{\rm c}(R=70)$
requires system sizes of at least $L=5000$, whereas we have carried out
simulations for system sizes up to $1000 \times 1000$ lattice sites. This has
been solved as follows. The renormalization treatment in Ref.~\cite{medran}
predicts the form of the function describing how $T_{\rm c}(R)$ deviates from
the mean-field critical temperature when $R$ varies. By fitting this function
to the accurately determined critical temperatures in our previous study an
expression is obtained for $T_{\rm c}(R)$ from which the critical temperatures
for very large ranges can be calculated to a relatively high accuracy.  The
shift of $T_{\rm c}$ is expressed by
\begin{equation}
 T_{\rm c} = T_{\rm c}^{\rm MF}
           + \frac{a_1}{R^2} \left[ 1 + a_2 \ln R^2 \right]
           + \frac{a_3}{R^4} \;,
\label{eq:tc-scale}
\end{equation}
where $T_{\rm c}^{\rm MF}=1$ and the last term is a higher-order correction
omitted in Ref.~\cite{medran}. A least-squares fit for $16 \lesssim R^2
\lesssim 70$ ($32 \leq R_m^2 \leq 140$ in Ref.~\cite{medran}) yielded
$a_1=-0.267 (6)$, $a_2=1.14 (3)$, and $a_3=-0.27 (3)$.
Figure~\ref{fig:kc-scale} shows the critical temperatures and
expression~(\ref{eq:tc-scale}) with the appropriate coefficients.

\subsection{Absolute magnetization density}
\begin{figure}
\begin{center}
\leavevmode
\epsfbox{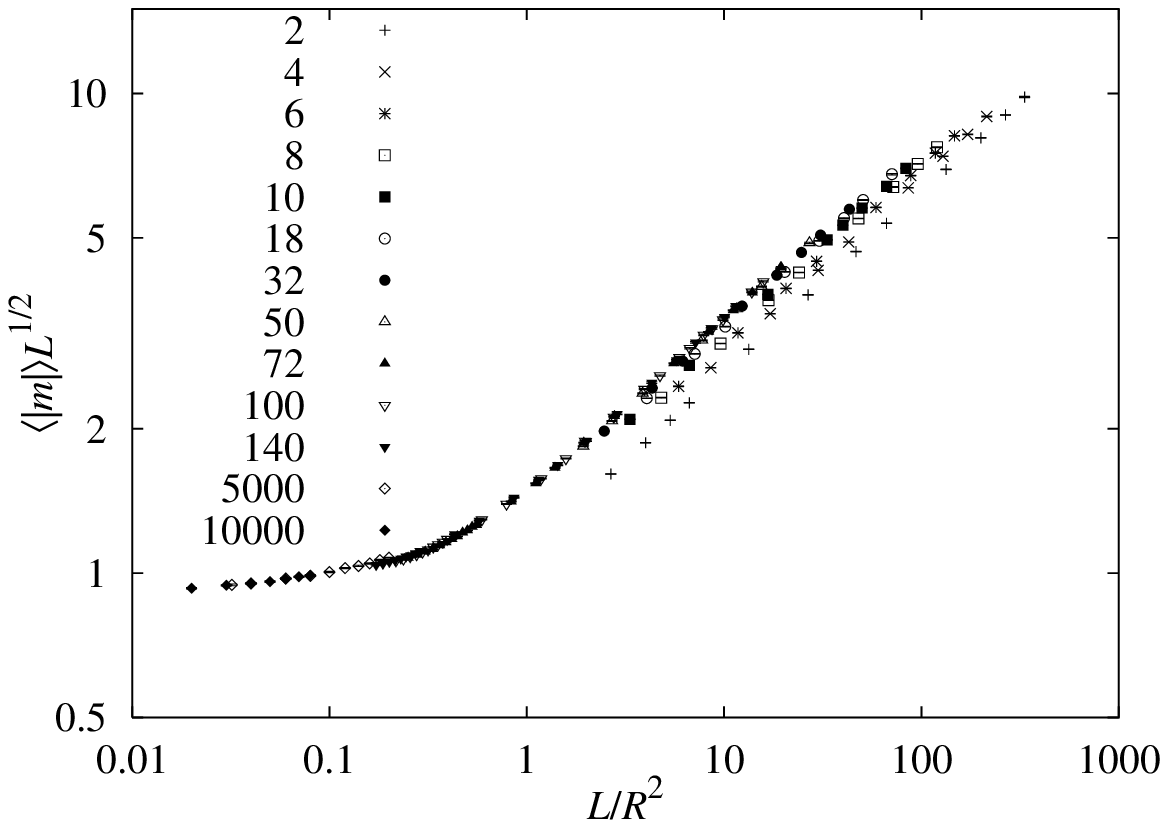} \\
\leavevmode (a) \\ \leavevmode
\epsfbox{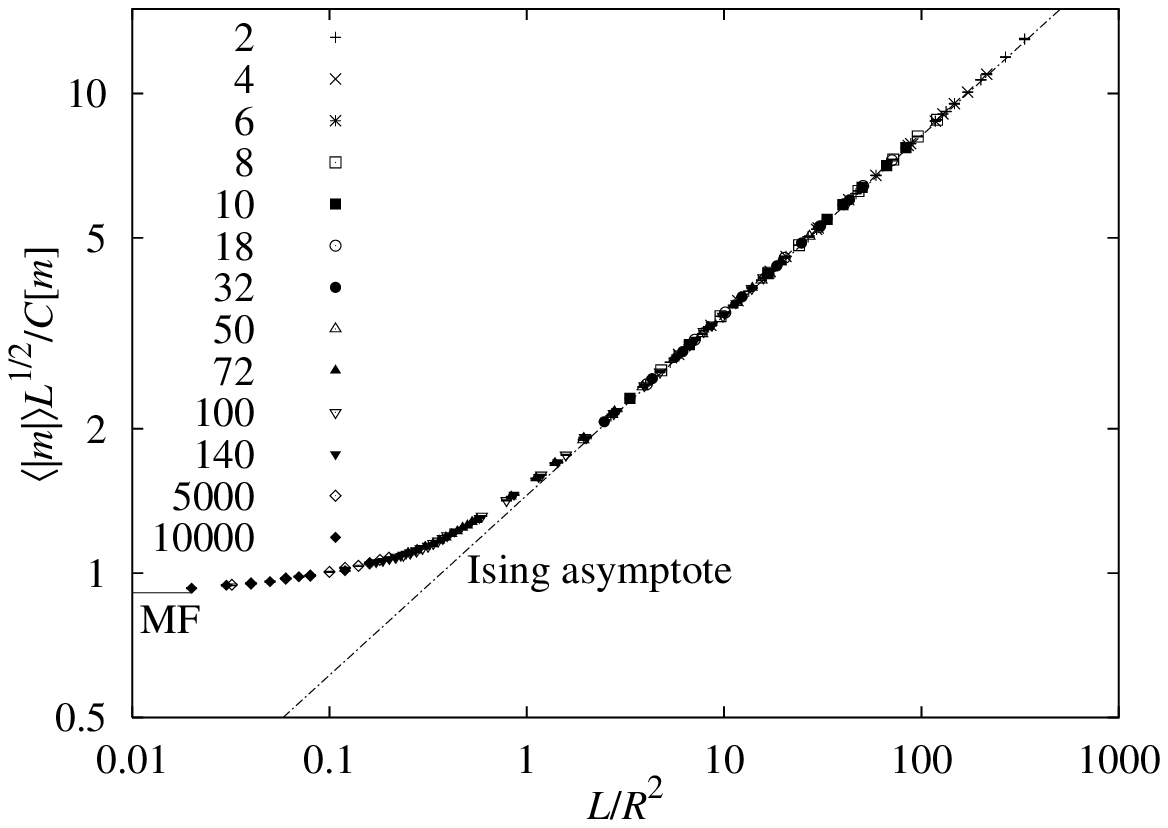} \\
\leavevmode (b)
\end{center}
\caption[]{(a)~Finite-size crossover curve for the absolute magnetization
  density multiplied by the square root of the system size. (b)~The same graph
  but now the range-dependent corrections predicted by renormalization theory
  have been divided out. The correction factor abbreviated by $C[m]$ stands for
  the factor between square brackets in Eq.~(\protect\ref{eq:magabs-scale}). A
  perfect collapse is obtained for all system sizes and interaction ranges.
  Both the exact mean-field limit (indicated by ``MF'') and the Ising asymptote
  with slope~$3/8$ are confirmed by the data. In this and all following figures
  the numbers in the key refer to values for $R_m^2$.}
\label{fig:m-cross}
\end{figure}

\begin{figure}
\begin{center}
\leavevmode
\epsfbox{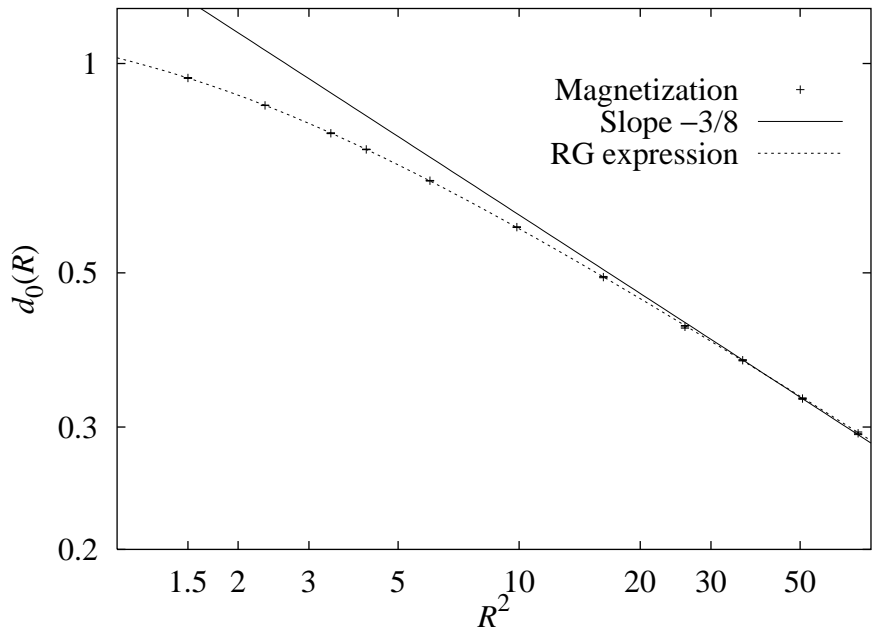}
\end{center}
\caption[]{Critical amplitude of $\langle |m| \rangle$ and the renormalization
  prediction fitted to it. This correction factor is used in
  Fig.~\protect\ref{fig:m-cross}(b).}
\label{fig:magabs}
\end{figure}

In the Ising regime, the absolute magnetization density scales (at criticality)
asymptotically as $\langle |m| \rangle = L^{-1/8}d_0(R)$, where the critical
amplitude $d_0$ is a function of $R$, $d_0 \propto R^{-3/4}$.  In the
mean-field regime $\langle |m| \rangle$ does not depend on $R$, but is simply
proportional to~$L^{-1/2}$. When plotting $\langle |m| \rangle$ as a function
of $G=L/R^2$ a data collapse is obtained if it is multiplied by a factor $L^x
R^{-(2x-1)}$. This resulting quantity is proportional to $G^{x-1/8}$ in the
Ising regime and to $G^{x-1/2}$ in the mean-field regime.  A suitable choice is
$x=1/2$, because this yields a quantity which is still independent of~$R$ in
the mean-field regime. Indeed, it is shown in the Appendix that in a
two-dimensional system in which all spin--spin interactions are equally strong,
\begin{equation}
\langle |m| \rangle =
    12^{1/4}\frac{\Gamma (\frac{1}{2})}{\Gamma (\frac{1}{4})}\frac{1}{\sqrt{L}}
    + {\cal O}\left(\frac{1}{L^{3/2}}\right)
\label{eq:mag-mf}
\end{equation}
and $\langle |m| \rangle \sqrt{L}$ will thus approach $12^{1/4} \Gamma(1/2) /
\Gamma(1/4) = 0.909890588\ldots$\ in the limit of $G \to 0$. Remark that our
requirement $L > \sqrt{2}R$ unambiguously relates the limit $G \to 0$ to the
mean-field ($R \to \infty$) limit.  In Fig.~\ref{fig:m-cross}(a) we have
plotted the absolute magnetization density multiplied by the square root of the
system size versus the crossover variable. Interaction ranges from $R_m^2=2$ to
$R_m^2=10000$ were included, where the data for $R_m^2=5000$ and $R_m^2=10000$
(spanning the range $0.02 \lesssim G \lesssim 0.2$) have been obtained at
temperatures calculated from Eq.~(\ref{eq:tc-scale}): $K_{\rm
  c}(R_m=\sqrt{5000})=6.3746(3) \times 10^{-5}$ and $K_{\rm
  c}(R_m=\sqrt{10000})=3.18491(9) \times 10^{-5}$. The crossover curve
evidently spans approximately three decades in~$G$. In the limit of $G \to 0$
it gradually approaches a horizontal line. For $G \gg 1$ the picture is not
very clear. The data points for each single value of $R$ lie on a straight line
with slope~$3/8$, corresponding to the Ising asymptote, but the asymptotes only
coincide for large ranges (cf.\ Fig.~4 in Ref.~\cite{mon}). The reason for this
is that, as mentioned above, for small ranges the critical amplitudes do not
conform to the leading $R^{-3/4}$~dependence. This can be cured by invoking the
renormalization treatment of Ref.~\cite{medran}. Indeed, the theory predicts
the structure of the corrections to the leading $R$~dependence of the critical
amplitude,
\begin{equation}
  d_0 = b_0 R^{-3/4} \left[ 1 + \frac{1}{R^2}(b_1 + b_2\ln R^2) \right] \;.
\label{eq:magabs-scale}
\end{equation}
This ``finite-range correction'' is very similar to the shift of the critical
temperature in Eq.~(\ref{eq:tc-scale}) but originates from a different term in
the renormalized LGW~Hamiltonian.  To illustrate this correction graphically,
we have reproduced Fig.~6 from Ref.~\cite{medran} and included the result of a
least-squares fit of Eq.~(\ref{eq:magabs-scale}) to the data, see
Fig.~\ref{fig:magabs}. The curve clearly yields an excellent description of the
critical amplitudes, even for small ranges. We have used this fit to construct
a clear crossover curve for the magnetization density on which the data for all
values of $R$ collapse. To this end, all data are divided by the correction
factor between square brackets in Eq.~(\ref{eq:magabs-scale}). The result is
shown in Fig.~\ref{fig:m-cross}(b).  One observes that in the Ising regime all
data perfectly collapse on a common asymptote with slope~$3/8$. For $G$ small
the data indeed approach the mean-field prediction~(\ref{eq:mag-mf}). The fact
that at $G \approx 0.2$ the data for $R_m^2=5000$ and $R_m^2=10000$ coincide
with those for $R_m^2=72$, $100$, $140$ confirms that the critical temperatures
for the large ranges have been estimated accurately. The center of the
crossover region lies between $G=0.1$ and $G=1.0$ which shows that the
parameter~$u$ is indeed of order unity. Finally, it is particularly encouraging
that no remaining finite-size effects, causing deviations from the curve, are
visible in Fig.~\ref{fig:m-cross}(b), despite the factor that the correction
factor was calculated in the $L \to \infty$ limit and hence does not compensate
for such higher-order finite-size effects.

\subsection{Magnetic susceptibility}
\label{sec:chi-fss}
\begin{figure}
\begin{center}
\leavevmode
\epsfbox{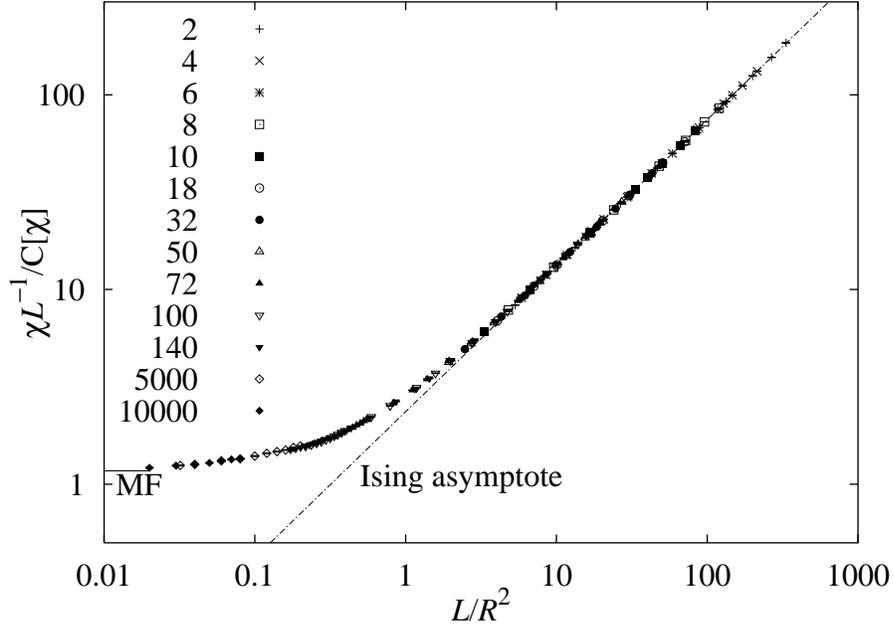}
\end{center}
\caption[]{Finite-size crossover curve for the magnetic susceptibility divided
  by the system size. The range-dependent correction factor~$C[\chi]$ [the
  factor between square brackets in Eq.~(\protect\ref{eq:chi-scale})] has been
  divided out, as discussed in the text. Both the mean-field limit and the
  Ising asymptote (slope~$3/4$) are confirmed by the data.}
\label{fig:chi-cross}
\end{figure}

The procedure described above for the absolute magnetization density can be
applied to the magnetic susceptibility~$\chi$, which we have calculated from
the average square magnetization, $\chi = L^d \langle m^2 \rangle$.  At
$T=T_{\rm c}$, the susceptibility is in the Ising regime proportional to
$L^{7/4}R^{-3/2}$, and in the mean-field regime it scales proportional to~$L$.
To obtain a data collapse for $\chi$ as a function of~$G$, one has to multiply
the finite-size data by $L^x R^{-(2x+2)}$, where a suitable choice is given by
$x=-1$. In the mean-field limit, $\chi/L$ approaches
$\sqrt{12}\Gamma(3/4)/\Gamma(1/4) = 1.17082866\ldots$\ (see Appendix). As shown
in Ref.~\cite{medran}, the deviation from the leading range dependence of the
critical amplitude is very similar to that of the absolute magnetization
density,
\begin{equation}
p_0 = q_0 R^{-3/2}
   \left[ 1 + \frac{1}{R^2}(q_1 + q_2 \ln R^2) + \frac{q_3}{R^4} \right] \;,
\label{eq:chi-scale}
\end{equation}
where now one additional higher-order correction is required.  Therefore we
only show the resulting crossover curve for the susceptibility after the data
have been divided by the correction factor between square brackets, see
Fig.~\ref{fig:chi-cross}. Again, both the mean-field asymptotic result and the
Ising asymptote (slope~$3/4$) are clearly reproduced, with a perfect collapse
for all ranges.

\subsection{Spin--spin correlation function}
\begin{figure}
\begin{center}
\leavevmode
\epsfbox{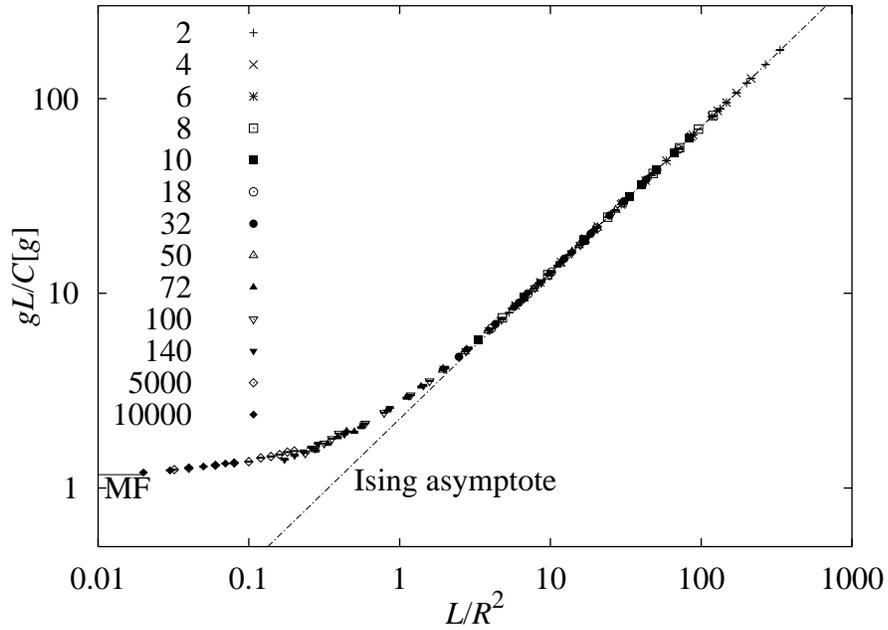}
\end{center}
\caption[]{Finite-size crossover curve for the spin--spin correlation function
  multiplied by the system size. A range-dependent correction factor
  (abbreviated as~$C[g]$) has been divided out, as discussed in the text. Both
  the mean-field limit and the Ising asymptote (slope~$3/4$) are confirmed by
  the data.}
\label{fig:g-cross}
\end{figure}

Closely related to the magnetic susceptibility is the spin--spin correlation
correlation function $g(\bfr)$. In our simulations we have sampled $g(L/2)$,
which scales both in the Ising regime and in the mean-field regime as
$\chi/L^2$. Thus, we obtain a data collapse by multiplying the finite-size data
by $L^x R^{-(2x-2)}$, in which we have set $x=1$. After correcting for the
higher-order range-dependent corrections in the critical amplitude [which have
the same structure as those in Eq.~(\ref{eq:chi-scale})] we obtain the graph
shown in Fig.~\ref{fig:g-cross}.  The full crossover curve can be mapped and
shows a close resemblance to that for the susceptibility, including the
approach of the asymptotic mean-field value.  In the range $0.2 \lesssim L/R^2
\lesssim 1.0$, the data do not precisely coincide on a smooth curve. This is
due to nonlinear finite-size effects, which are for the spin--spin correlation
function apparently larger than for the absolute magnetization density or the
magnetic susceptibility. We will pay more attention to these deviations when
discussing the universal amplitude ratio (see below). It should be noted that
the critical amplitudes listed in Table~V of Ref.~\cite{medran} have to be
multiplied by a factor $2^{-1/4} = 0.84089642\ldots$\ in order to obtain the
correct values.

\subsection{Universal amplitude ratio}
\begin{figure}
\begin{center}
\leavevmode
\epsfbox{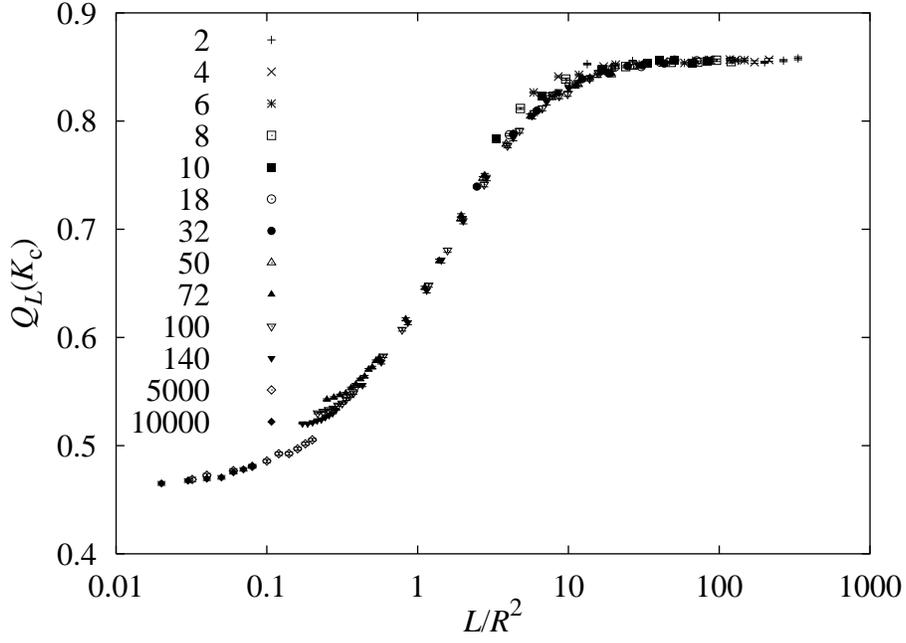} \\
\leavevmode (a) \\ \leavevmode
\epsfbox{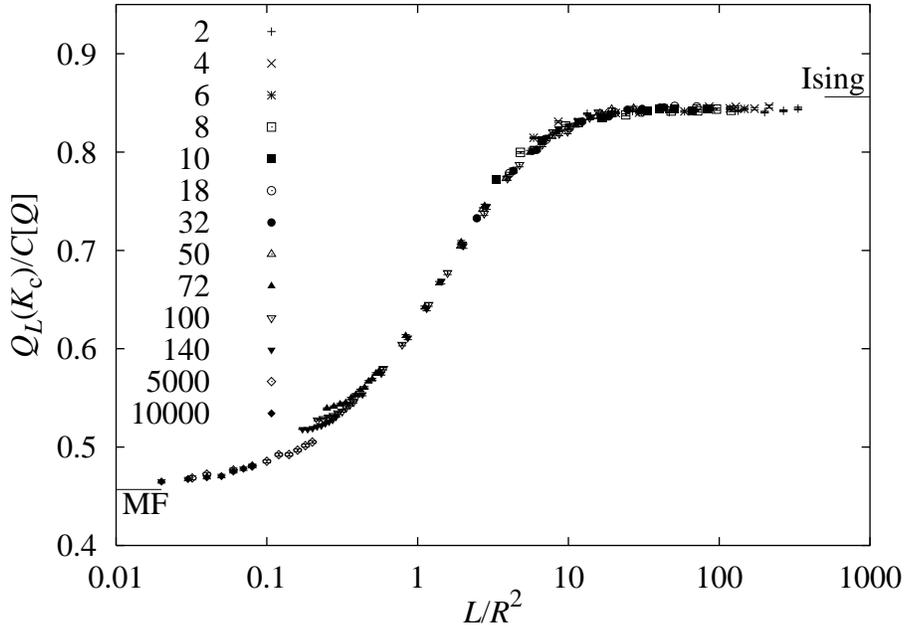} \\
\leavevmode (b)
\end{center}
\caption[]{Finite-size crossover curves for the amplitude ratio $Q \equiv
  \langle m^2 \rangle^2 / \langle m^4 \rangle$. Figure~(a) shows the curve
  without any additional corrections, whereas in~(b) a range-dependent
  correction factor~$C[Q]$ [see Eq.~(\protect\ref{eq:qfactor})] has been
  divided out. For small values of the crossover variable~$L/R^2$ the
  mean-field limit is reproduced and for large values of~$L/R^2$ the Ising
  limit is approached. For a further discussion see the text.}
\label{fig:q-cross}
\end{figure}

\begin{figure}
\begin{center}
\leavevmode
\epsfbox{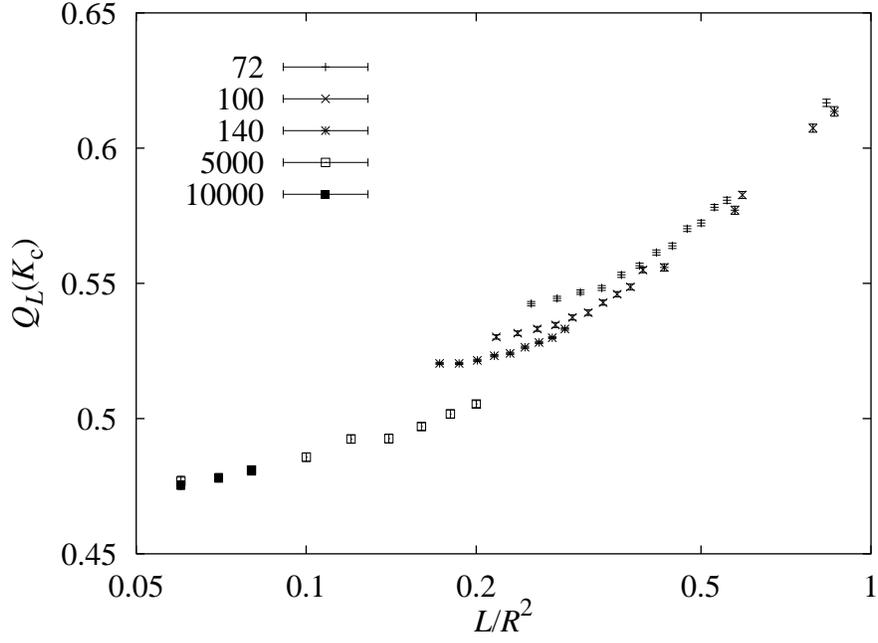}
\end{center}
\caption[]{A detailed view of Fig.~\protect\ref{fig:q-cross}(a) showing the
  deviations from the crossover curve for very small system sizes.}
\label{fig:q-zoom}
\end{figure}

\begin{figure}
\begin{center}
\leavevmode
\epsfbox{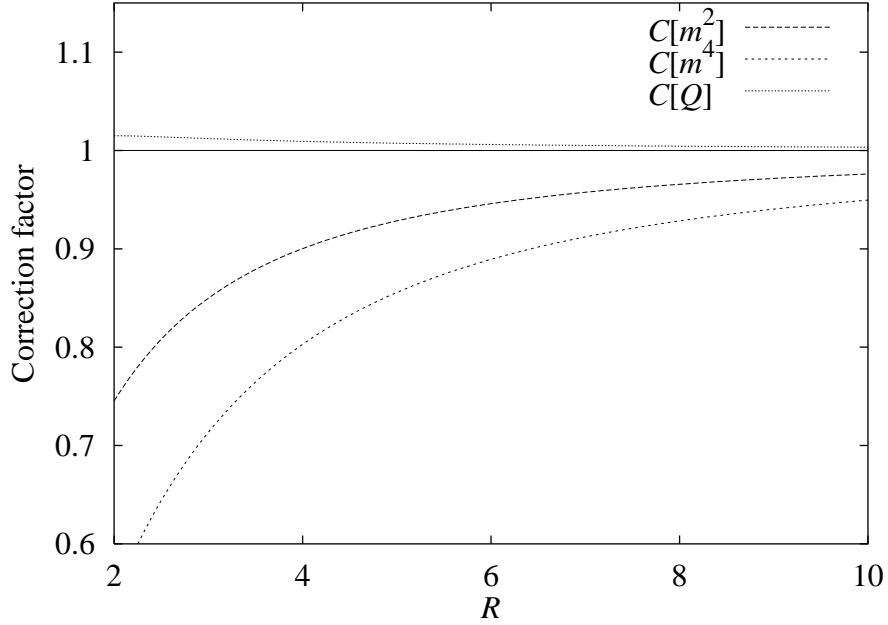}
\end{center}
\caption[]{The range-dependent correction factors $C[m^2] = C[\chi]$, $C[m^4]$,
  and~$C[Q]$ in $\langle m^2\rangle$, $\langle m^4 \rangle$, and~$Q$,
  respectively, as determined by least-squares fits to the critical amplitudes
  extracted from the Monte Carlo data. The line at height~$1$ is drawn for
  reference. One observes that $C[Q]$ lies very close to, although not exactly
  at, unity.}
\label{fig:corrfac}
\end{figure}

The amplitude ratio $Q_L \equiv \langle m_L^2 \rangle^2 / \langle m_L^4
\rangle$ is a size-dependent quantity, which takes a universal value~$Q$ in the
$L \to \infty$ limit. That is, it is calculated by taking the ratio of the
square of the magnetization density and the fourth power of it in a finite
geometry and subsequently taking the limit $L \to \infty$.  For $T>T_{\rm c}$,
$Q$ approaches the Gaussian value~$Q=1/3$ and for $T<T_{\rm c}$ it approaches
the maximum value~$Q=1$. At criticality, the amplitude ratio is known exactly
in the mean-field case, $Q_{\rm MF}=0.45694658\ldots$~\cite{bzj,ijmpc} and to a
high accuracy in the two-dimensional Ising model, $Q_{\rm I} \approx
0.856216(1)$~\cite{kamblo}. In Ref.~\cite{medran}, $Q_L(K_{\rm c})$ was plotted
for a large interaction range ($R_m^2=140$) as a function of the system size.
The approach of the Ising value was clearly visible for $L$ large, but for
small system sizes $Q$ first decreased towards $Q_{\rm MF}$ and then started to
show strong finite-size effects. Evidently, it is a better approach to
construct the true crossover curve for $Q(K_{\rm c})$ by plotting finite-size
data for $Q$ for various ranges versus the crossover variable. This is shown in
Fig.~\ref{fig:q-cross}(a). Several remarks apply to this graph. Firstly, one
notes that $L/R^2$ is indeed the appropriate crossover variable: a reasonable
collapse is obtained for all values of $L$ and~$R$. However, some remarkable
deviations from this scaling behavior are present, which are most clearly
visible in the range $0.2 < L/R^2 < 0.6$, but also present around $L/R^2=10$.
Similar effects were already observed in the spin--spin correlation function,
but now the effects stand out much more pronounced because we have employed for
the amplitude ratio a linear instead of a logarithmic vertical scale.  These
deviations are due to nonlinear finite-size corrections, as can be seen clearly
by zooming in into the deviations, see Fig.~\ref{fig:q-zoom}. The data points
for $R_m^2=5000$ and $R_m^2=10000$ may serve as a reference for the location of
the ``true'' crossover curve. One observes that for each of the ranges
$R_m^2=72$, $100$, and~$140$ the deviations from this curve increase with {\em
  decreasing\/} system size, which indeed shows that the effects are caused by
finite-size corrections. If the deviations had been caused by, e.g., an
inaccurate determination of the critical temperature, the effects would have
increased with increasing system size. Unfortunately, it is not easy to
separate these corrections from the leading crossover behavior (except
graphically), unless the full crossover function is known (which in turn would
limit the use of a numerical determination). Of course this problem can be
circumvented by determining the crossover at these values for~$G$ from systems
with a larger system size and a larger interaction range. The deviations around
$L/R^2=10$ are caused by the same effect, but now for systems with small~$R$.
Although the amplitude ratio is more sensitive---even if one takes into account
the difference in scale---to these finite-size effects than $\langle m^2
\rangle = \chi/L^2$ and~$\langle m^4 \rangle$ individually (the curve for the
latter is not shown here, but its smoothness is comparable to that of the
susceptibility), $Q$ is less sensitive to corrections to the leading range
dependence. Indeed, for $\langle m^4 \rangle$ these corrections are again of
the form $[1 + R^{-2}(s_1 + s_2 \ln R^2) + R^{-4}s_3]$ and $Q$ must thus be
divided by
\begin{equation}
\frac{[1 + R^{-2}(q_1 + q_2 \ln R^2) + R^{-4}q_3]^2}%
     {1 + R^{-2}(s_1 + s_2 \ln R^2) + R^{-4}s_3} \;.
\label{eq:qfactor}
\end{equation}
The coefficients $s_1$, $s_2$, and~$s_3$ have been determined from a
least-squares fit to the critical amplitudes of $\langle m^4 \rangle$ and
$q_1$, $q_2$, and~$q_3$ come from Eq.~(\ref{eq:chi-scale}).
Figure~\ref{fig:corrfac} shows the correction factors for $\langle m^2
\rangle$, $\langle m^4 \rangle$, and $Q$.  Evidently, the latter
factor~(\ref{eq:qfactor}) is much closer to unity than the former two.
Figure~\ref{fig:q-cross}(b) shows $Q_L(K_{\rm c})$ divided by the correction
factor~(\ref{eq:qfactor}), which indeed shows only slightly less scatter than
the graph without this correction factor. In particular the deviations for the
larger ranges do not disappear.

\section{Thermal crossover scaling}
\label{sec:thermal}

\subsection{General considerations}
The finite-size crossover scaling studied in the previous section is an
intrinsic finite-size effect which is not observable in thermodynamic systems.
For this reason it is important to study its temperature-dependent counterpart
as well. This so-called {\em thermal\/} crossover, which was from a
phenomenological scaling point of view already considered in
Ref.~\cite{riedel}, is of course closely related to finite-size crossover: in
finite systems crossover to mean-field-like behavior occurs when the {\em
  system size\/} has been decreased to the appropriate power of the interaction
range (i.e.\ $L \sim R^{4/(4-d)}$ or $L \sim R^2$ for $d=2$), whereas in the
temperature-dependent case this crossover occurs when the temperature distance
to the critical point is such the {\em correlation length\/} has become of the
order of an appropriate power of the interaction range. In the latter case, the
precise crossover location is determined by the Ginzburg criterion,
$t^{(4-d)/2}R^d u^{-1} \approx 1$, where $u$ is the coefficient of the
$\phi^4$~term in the LGW~Hamiltonian. It should be kept in mind that these
considerations are valid only {\em within the critical region}, i.e.\ care must
be exercised to keep the reduced temperature sufficiently small. When studying
thermal crossover in practical simulations one has the additional complication
that sufficiently close to $T_{\rm c}$ the correlation length will always be
bounded by the finite system size, which is precisely the situation one wants
to avoid. So relatively large system sizes are required.

As follows from the Ginzburg criterion, the appropriate scaling variable in two
dimensions is $tR^2$ and one can therefore study thermal crossover effects by
varying the interaction range as well. This is essential because of the
following. For small values of~$R$, $t$ has to be made rather large to cross
over to classical critical behavior and it is possible that one leaves the
critical region before reaching the classical regime.  On the other hand, if
one only studies systems with large interaction ranges, $t$ has to be made very
small to observe Ising-like critical behavior. However, for such small values
of~$t$ extremely large system sizes are required to avoid finite-size effects.
Therefore we have constructed, just as in the previous section, crossover
curves from results for various ranges.  We have carried out simulations for
the interaction ranges studied in Ref.~\cite{medran} at temperatures further
below~$T_{\rm c}$ and also generated data for the interaction ranges
$R_m^2=500$, $1000$, $4000$, and~$10000$.  Table~\ref{tab:largerange}
summarizes some properties of these systems. Simulations have been carried out
down to temperatures as low as $T \approx 0.5T_{\rm c}$.  For the order
parameter crossover can only be studied in the phase of broken symmetry, but
for the susceptibility we have also considered the symmetric ($T>T_{\rm c}$)
phase.  Since in this phase no saturation effects occur, much smaller
interaction ranges suffice to span the full crossover region, as we will show
below.

\subsection{Absolute magnetization density}
\label{sec:mag-therm}
\begin{figure}
\begin{center}
\leavevmode
\epsfbox{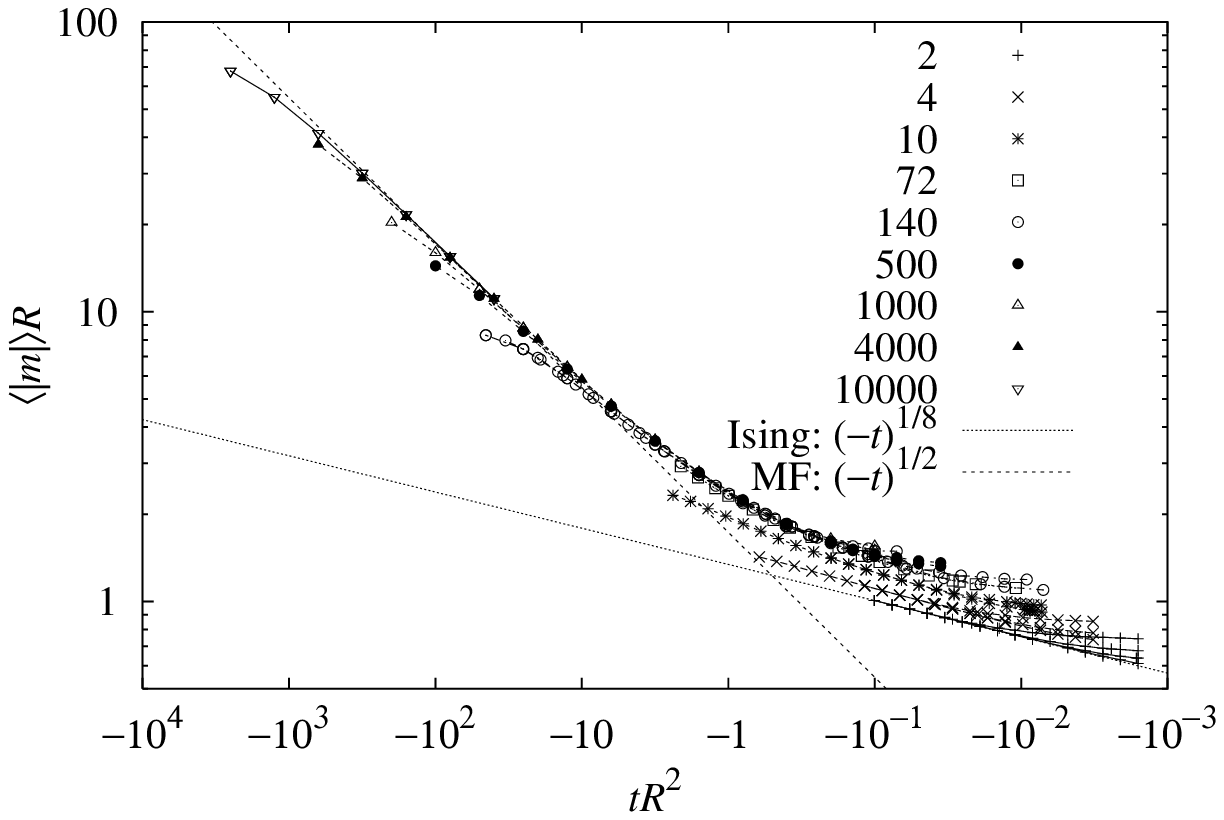} \\
\leavevmode (a) \\ \leavevmode
\epsfbox{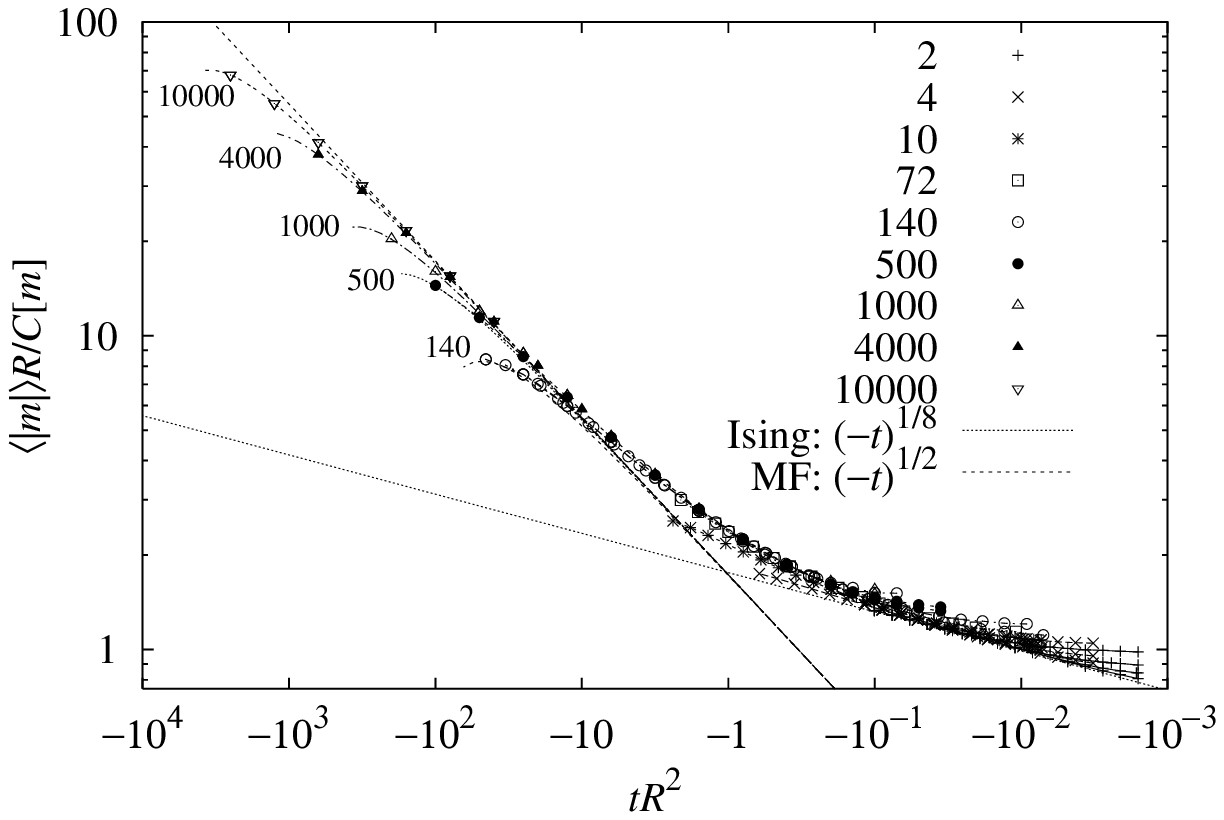} \\
\leavevmode (b) \\ \leavevmode
\epsfbox{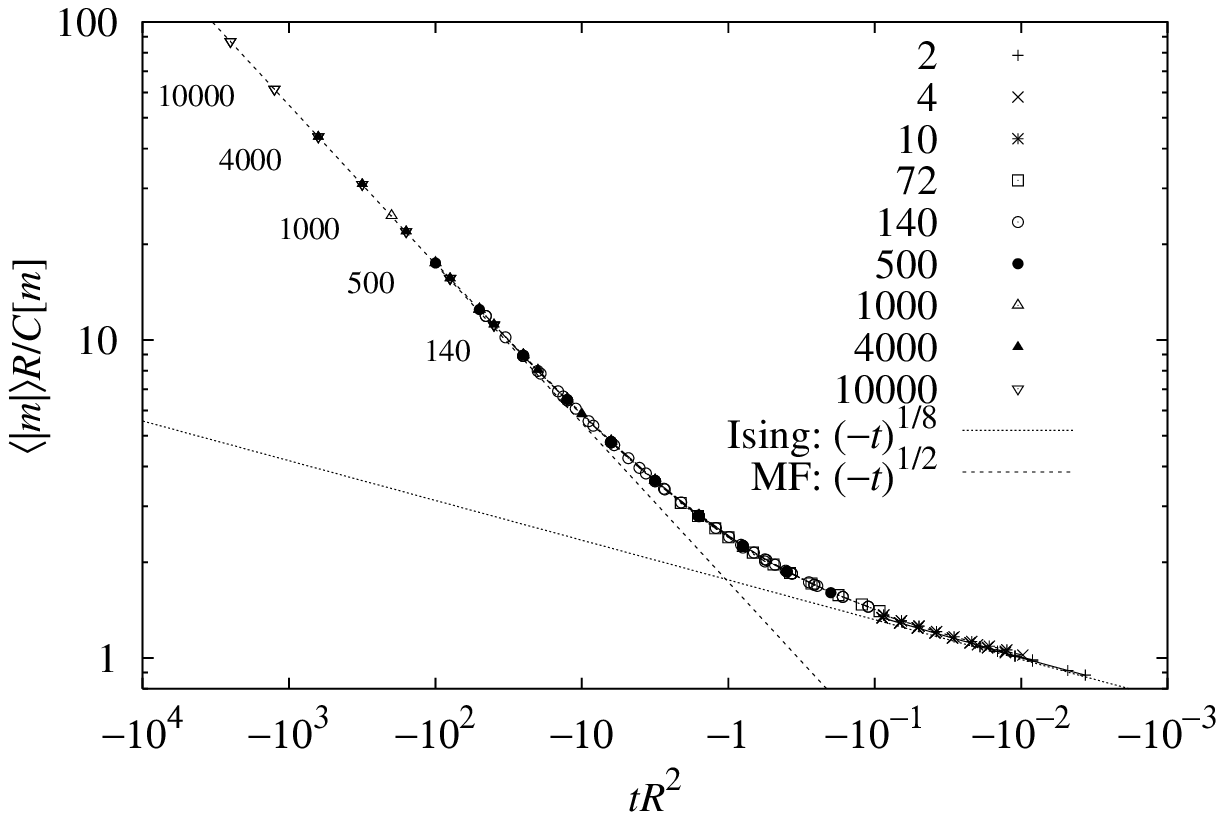} \\
\leavevmode (c)
\end{center}
\caption[]{Thermal crossover for the absolute magnetization density for various
  ranges and system sizes. In figure~(a) no additional correction terms have
  been used, whereas in~(b) the factor~$C[m]$ has been divided out.  In
  figure~(c) the data for $R_m^2 \geq 72$ have also been corrected for
  saturation effects and data points in the finite-size regime have been
  omitted. For an extensive discussion of the various features of these graphs
  the reader is referred to the text.}
\label{fig:mag-temp}
\end{figure}

As derived in Refs.~\cite{mon,medran} the absolute magnetization density
scales, sufficiently close to the critical point, as $\langle |m| \rangle
\propto (-t)^{\beta}R^{(2d\beta-d)/(4-d)}$ ($t<0$), which for the
two-dimensional case yields $\langle |m| \rangle \propto (-t)^{1/8}R^{-3/4}$.
In the mean-field regime, on the other hand, the magnetization density is
simply proportional to $(-t)^{1/2}$. When plotted as a function of $tR^2$, a
data collapse for all ranges is now obtained if the magnetization density is
multiplied by $R$.  Figure~\ref{fig:mag-temp}(a) shows the corresponding plot.
We will discuss the various aspects of this graph in some more detail.  The
overall picture suggests that the data roughly follow the Ising asymptote
(slope~$1/8$) for small values of $tR^2$ and then gradually approach the
mean-field asymptote (slope~$1/2$) for large values of $tR^2$.  Here ``small''
and ``large'' refer to the absolute value of $tR^2$ and ``slope'' is generally
used for the logarithmic derivative, $d \ln \langle |m| \rangle / d \ln |t|$.
For very small values of $tR^2$ the data start to deviate from the Ising
asymptote at an $L$-dependent location and approximately follow (for
temperatures closer to $T_{\rm c}$) a horizontal line. Here one has entered the
finite-size regime, where the correlation length is limited by the system size.
This is the case which was studied in the previous section.  The width of this
regime depends (for general~$d$) both on the system size and the interaction
range, as can be read off from the universal scaling functions derived in
Ref.~\cite{medran}. Indeed, the temperature-dependent argument of these
functions is $tL^{y_{\rm t}} R^{-2(2y_{\rm t}-d)/(4-d)}$ ($y_{\rm t}=1$ in the
2D~Ising universality class) and the width of the finite-size regime is thus
proportional to $L^{-y_{\rm t}} R^{2(2y_{\rm t}-d)/(4-d)} = L^{-1}$.  Note that
the absence of any range dependence is {\em not\/} a general feature and even
for the two-dimensional Ising model only true to leading order (cf.\ Fig.~5 of
Ref.~\cite{medran}).  Higher order terms will entail range-dependent factors
that involve (for $d=2$) logarithms of~$R$. Outside of the finite-size regime,
the data for each individual range first lie approximately on the Ising
asymptote, which has been drawn with an amplitude such that it coincides with
the data for $R_m^2=2$. For the smaller ranges the amplitudes of the asymptotes
show a considerable range dependence, whereas for larger ranges the amplitudes
converge. Upon further decrease of the temperature (increase of the absolute
value of~$t$) several types of behavior occur: for the smallest range
($R_m^2=2$) the data points still lie on the Ising asymptote.  For $R_m^2=4$
and $R_m^2=10$ the data leave the Ising asymptote at sufficiently low
temperatures and then follow a nearly straight line with a slope that lies
between the Ising and the mean-field asymptote. In these cases one has left the
critical region without ever reaching the asymptotic mean-field regime. For
each range the data for all system sizes coincide, as they should outside of
the finite-size regime.  For $R_m^2=72$ and $R_m^2=140$ the mean-field
asymptote is approached much closer.  However, if the temperature is decreased
further below the critical temperature the data points start to deviate from
the asymptote again. This effect is caused by saturation of the magnetization
and can be quantitatively described with mean-field theory, as we will show
below. Turning to even larger ranges, we see that the data now really reach the
asymptote with slope~$1/2$ and follow it for up to one decade in the crossover
variable (for the largest range we have studied) before saturation sets in.
Also the exact amplitude~$\sqrt{3}$ (see below) of the asymptote is precisely
reproduced, which shows again that the critical temperatures of the systems
with large interaction ranges have been accurately determined: a deviation
would have shifted the graph along the horizontal axis.

We will now first consider the offset of the asymptotes in the Ising regime.
Although this effect occurs outside the finite-size regime, we may well hope
that the so-called finite-range corrections applied in the previous section
[Eq.~(\ref{eq:magabs-scale})] can be used here as well. Indeed, these
corrections are part of the universal scaling functions and although the
amplitude $b_0 = \lim_{R \to \infty} \lim_{L \to \infty} R^{3/4} L^{1/8}
\langle |m_L(K_{\rm c})| \rangle$ is a specific limiting value, the
range-dependent correction factor does not depend on this limit.  Especially
the collapse obtained in Fig.~\ref{fig:m-cross}(b) makes it very tempting to
apply a similar correction here.  On the other hand, these corrections were
calculated in the {\em Ising\/} regime, which we here are gradually leaving.
In Fig.~\ref{fig:mag-temp}(b) we show the same data but now divided by the
correction factor. Although a perfect collapse is not obtained, the asymptotes
lie much closer together than without this correction.

Also the critical amplitude of the Ising asymptote is known exactly. Indeed,
by expanding Onsager's expression for the spontaneous
magnetization~\cite{onsager49,mccoy_wu},
\begin{equation}
m = \left[ 1 - \frac{1}{\sinh^4(2J/k_{\rm B}T)} \right]^{1/8} \;,
\end{equation}
around the critical point $J/k_{\rm B}T_{\rm c} =
\frac{1}{2}\arcsinh(1)$ we obtain for $t<0$
\begin{equation}
m = [4\sqrt{2}\arcsinh(1) (-t) + {\cal O}(t^2)]^{1/8}
    \approx 1.22240995 (-t)^{1/8} \;.
\label{eq:ising-asymptote}
\end{equation}
For the nearest-neighbor Ising model $R=R_m=1$, so the fact that in
Fig.~\ref{fig:mag-temp} along the horizontal axis $tR^2$ is used instead of~$t$
and along the vertical axis $\langle |m| \rangle R$ instead of~$\langle |m|
\rangle$ does not affect the amplitude of the asymptote. However, the
correction factor~$C[m]$ [denoting the factor between square brackets in
Eq.~(\ref{eq:magabs-scale})] must of course be taken into account.  This
correction factor describes the deviation of the critical amplitude~$d_0(R)$
from the leading scaling behavior in terms of a power series in~$R^{-2}$ (with
coefficients that depend on~$\ln R$) and it is not {\em a priori\/} clear
whether $C[m]$ converges for $R=1$. It is certainly unlikely that a single term
[the term proportional to~$b_2$ in~(\ref{eq:magabs-scale}) vanishes] describes
the deviation very well. No exact result for $d_0(R=1) = \lim_{L \to \infty}
m_L(K_{\rm c})L^{1/8}$ is known to us, but from a modest Monte Carlo simulation
we found $d_0(R=1)=1.0092(4)$. On the other hand, from
Eq.~(\ref{eq:magabs-scale}) with $b_0=1.466(2)$ and $b_1=-0.305(1)$ we find
$d_0(R=1) = 1.018(4)$ which differs approximately two standard deviations from
the numerical result. Recall that $b_0$ and~$b_1$ were obtained from a
least-squares fit to the critical finite-size amplitudes for $2 \leq R_m^2 \leq
140$.  Nevertheless, the relative difference lies below the one-percent level,
which cannot be distinguished in our graph.  Therefore we have drawn the Ising
asymptote with amplitude $[4\sqrt{2}\arcsinh(1)]^{1/8}/(1-b_1)$ in
Fig.~\ref{fig:mag-temp}(b) and it indeed turns out to be a precise tangent to
the crossover curve.

As mentioned above, also the saturation effects can be described with
mean-field theory. Namely, the magnetization follows from the well-known
expression~\cite{bragg,baxter}
\begin{equation}
 m = \tanh \left( \frac{T_{\rm c}}{T}m \right) \;.
\end{equation}
Rewriting this as $m=(1+t)\arctanh(m)$ and solving for $m$ one obtains
below $T_{\rm c}$ for small~$t$
\begin{equation}
m = \sqrt{3}(-t)^{1/2} - \frac{2}{5}\sqrt{3}(-t)^{3/2}
  - \frac{12}{175}\sqrt{3}(-t)^{5/2} - \frac{2}{125}\sqrt{3}(-t)^{7/2}
  + \frac{166}{67375}\sqrt{3}(-t)^{9/2} + {\cal O}((-t)^{11/2}) \;.
\label{eq:magseries}
\end{equation}
The leading term shows the classical value $\beta=1/2$ and the critical
amplitude~$\sqrt{3}$. To describe the saturation effects in
Fig.~\ref{fig:mag-temp}, the first three terms of this series suffice.
Figure~\ref{fig:mag-temp}(b) shows for the five largest ranges ($R_m^2=140$,
$500$, $1000$, $4000$, $10000$) the curves
\begin{equation}
\langle |m| \rangle R = \sqrt{3}(-tR^2)^{1/2}
  \left[1 - \frac{2}{5 R^2}(-tR^2) - \frac{12}{175 R^4}(-tR^2)^2 \right] \;.
\label{eq:magseries-r}
\end{equation}
For $R_m^2=140$ this expression does not precisely coincide with the numerical
data, but for the remaining values the curves accurately describe the
saturation effects. For these cases the interaction ranges are apparently large
enough to suppress the critical fluctuations to a large extent. The lowest
temperatures shown in the figure are $T/T_{\rm c}=0.52$, $0.60$, $0.60$,
$0.68$, and~$0.50$ for $R_m^2=140$, $500$, $1000$, $4000$, and~$10000$,
respectively. Saturation effects become visible in Fig.~\ref{fig:mag-temp} for
$t \lesssim -0.15$, i.e.\ $T/T_{\rm c} \lesssim 0.85$. According to
Eq.~(\ref{eq:magseries}) the magnetization deviates here approximately five
percent from the asymptote.  Using Eq.~(\ref{eq:magseries}) we can perform
another operation on the numerical data. Namely, the influence of saturation
effects in the mean-field model is described by the ratio between the full
series expansion on the right-hand side of~(\ref{eq:magseries}) and its first
term. As the mean-field expression constitutes an accurate description of the
saturation effects for $R_m^2 \geq 500$, the factor between square brackets in
Eq.~(\ref{eq:magseries-r}) will give an accurate description of the {\em
relative\/} saturation effects (i.e., the ratio of the saturated magnetization
and the crossover curve) down to probably even lower interaction ranges. To
illustrate this we have divided the data for $R_m^2 \geq 72$ by the
corresponding factor. The resulting graph [Fig.~\ref{fig:mag-temp}(c)]---in
which also the data points in the finite-size regime have been omitted---shows
that the data for these large ranges now nicely coincide on one curve, which is
the actual crossover curve for the order parameter.

The fact that for different interaction ranges the data (which overlap for
considerable intervals of~$tR^2$) coincide on one curve lends strong support to
the hypothesis that the crossover curve is universal. As has already been noted
in Ref.~\cite{chicross}, this contrasts with the conclusion drawn by Anisimov
{\em et al.}~\cite{anisimov92} for the three-dimensional case. In
Ref.~\cite{anisimov92} it has been suggested that in the crossover region
microscopic cutoff effects are not negligibly small compared to the {\em
  finite\/} correlation length~$\xi$, which implies that the form of the
crossover curve depends on the ratio between~$\xi$ and the lattice spacing~$a$.
In our simulations we have not measured the correlation length directly, but we
can still make a rough estimate from the data. Namely, at the locations marking
the boundaries of the finite-size regime for different interaction ranges and
system sizes in Fig.~\ref{fig:mag-temp} the correlation length is approximately
equal to the system size. From the magnetization densities for $R_m^2 \geq 72$
we conclude that $\xi \approx 0.5/(-t)$, independent of the interaction range.
The latter conclusion is in agreement with the above-mentioned renormalization
prediction that the width of the finite-size regime is to leading order
independent of the interaction range. Thus, at a fixed value of the crossover
variable~$tR^2$ the correlation lengths for different ranges have {\em
  different\/} values. However, the crossover curves coincide at fixed~$tR^2$
and hence are independent of the ratio~$\xi/a$.

Finally, we make some observations concerning the size of the crossover region.
It is clear that it takes between two and three decades in the crossover
variable to cross over from Ising-like to classical critical behavior. Thus,
unless one studies systems with a rather large interaction range, one has to go
to such a large temperature distance from $T_{\rm c}$ to sufficiently decrease
the correlation length compared to the interaction range that one has already
left the critical region before observing classical critical behavior! The
center of the crossover region lies in the neighborhood of~$|tR^2|=1$,
consistent with a value for~$u$ of order unity.

\subsection{Magnetic susceptibility}
\begin{figure}
\begin{center}
\leavevmode
\epsfbox{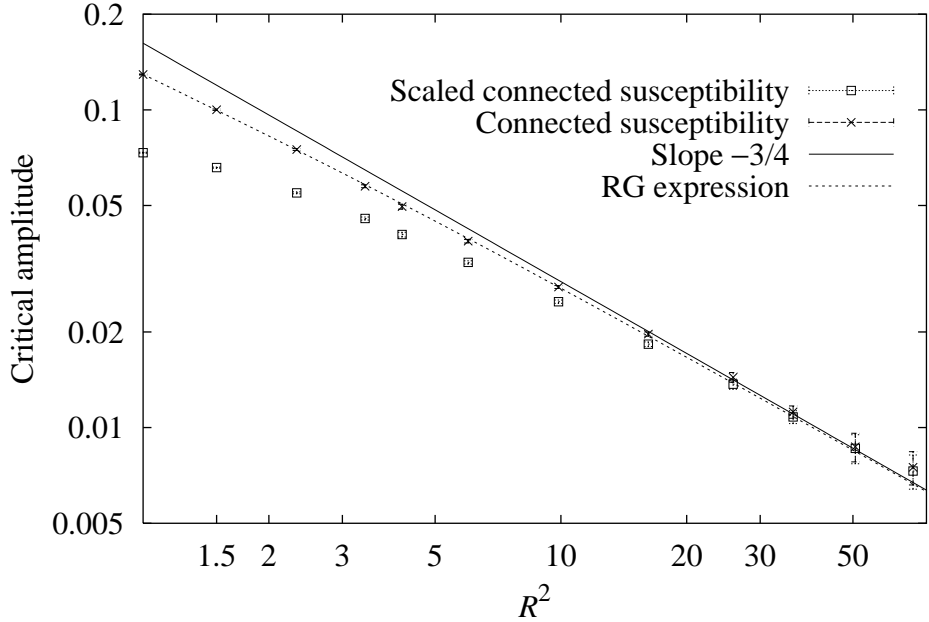}
\end{center}
\caption[]{Critical amplitude for the connected susceptibility $\tilde{\chi} =
  L^d (\langle m^2 \rangle - \langle |m| \rangle^2)/k_{\rm B}T$ as extracted
  from the thermodynamic limit of $L^{-7/4}\tilde{\chi}_L(K_{\rm c})$. The
  dashed curve indicates the renormalization prediction fitted to the numerical
  data.  Also the critical amplitude of the scaled susceptibility $k_{\rm
  B}T_{\rm c}\tilde{\chi}$ is shown, which for small ranges deviates
  considerably stronger from the asymptotic behavior.}
\label{fig:con-amp}
\end{figure}

Unlike the order parameter, the magnetic susceptibility displays crossover upon
approaching the critical point either from below or from above.  We will
discuss these two situations separately. In the ordered phase, $T \leq T_{\rm
  c}$, we approximate the magnetic susceptibility by the so-called connected
susceptibility,
\begin{equation}
\tilde{\chi} =
   L^d \frac{\langle m^2 \rangle - \langle |m| \rangle^2}{k_{\rm B}T} \;.
\end{equation}
In the two-dimensional Ising model with interaction range~$R$ this quantity
will, close to the critical point, diverge as $(-t)^{-7/4}R^{-3/2}$. Further
below~$T_{\rm c}$ it will cross over to classical critical behavior, where
$\tilde{\chi} \propto (-t)^{-1}$. In a graph showing results for various ranges
as a function of the crossover variable~$tR^2$ a data collapse is obtained for
$\tilde{\chi}/R^2$. However, just as for previous crossover curves, the data
for small~$R$ will display an offset because of corrections to the leading
$R^{-3/2}$~dependence. To determine these deviations we first study the
critical amplitude of the connected susceptibility, which was not considered in
Ref.~\cite{medran}; see Fig.~\ref{fig:con-amp}. The statistical uncertainty of
this amplitude is notably larger than for $\langle |m| \rangle$ and~$\langle
m^2 \rangle$ (cf., e.g., Fig.~\ref{fig:magabs}), but one can still observe that
the asymptotic regime is reached. In this figure we have also plotted the
critical amplitude of the so-called scaled susceptibility~$k_{\rm
  B}T\tilde{\chi}$ which was studied in, e.g., Ref.~\cite{mon}.  Evidently, the
latter amplitude shows a much stronger deviation from the leading range
dependence, due to the fact that also $T_{\rm c}(R)$ deviates from $T_{\rm
  c}^{\rm MF}$ (Fig.~\ref{fig:kc-scale}). Thus, although both amplitudes have
the same asymptotic behavior for large interaction ranges, it is much more
difficult to extract this behavior from medium-range results for $k_{\rm
  B}T\tilde{\chi}$ than from the corresponding results for~$\tilde{\chi}$. This
may partially explain the difficulties experienced in Ref.~\cite{mon}.  The
deviations have been fitted to a correction factor of the form $[1 + R^{-2}(v_1
+ v_2 \ln R^2)]$, which we abbreviate as $C[\tilde{\chi}]$. Indeed, the fact
that the finite-range corrections for~$\tilde{\chi}$ are so small allowed us to
neglect them altogether in Ref.~\cite{chicross}, where only logarithmic scales
have been employed.

\begin{figure}
\begin{center}
\leavevmode
\epsfbox{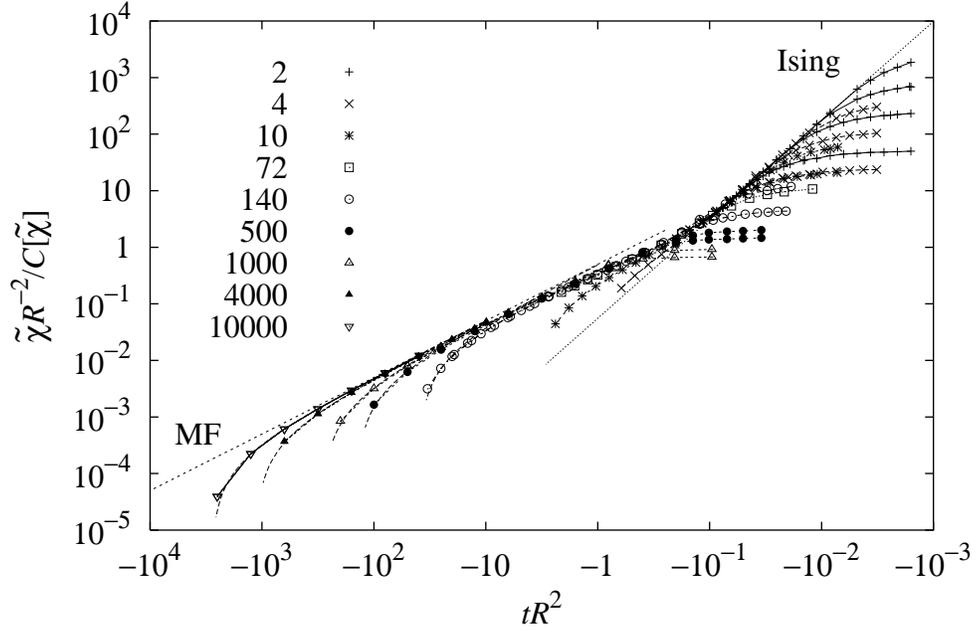} \\
\leavevmode (a) \\ \leavevmode
\epsfbox{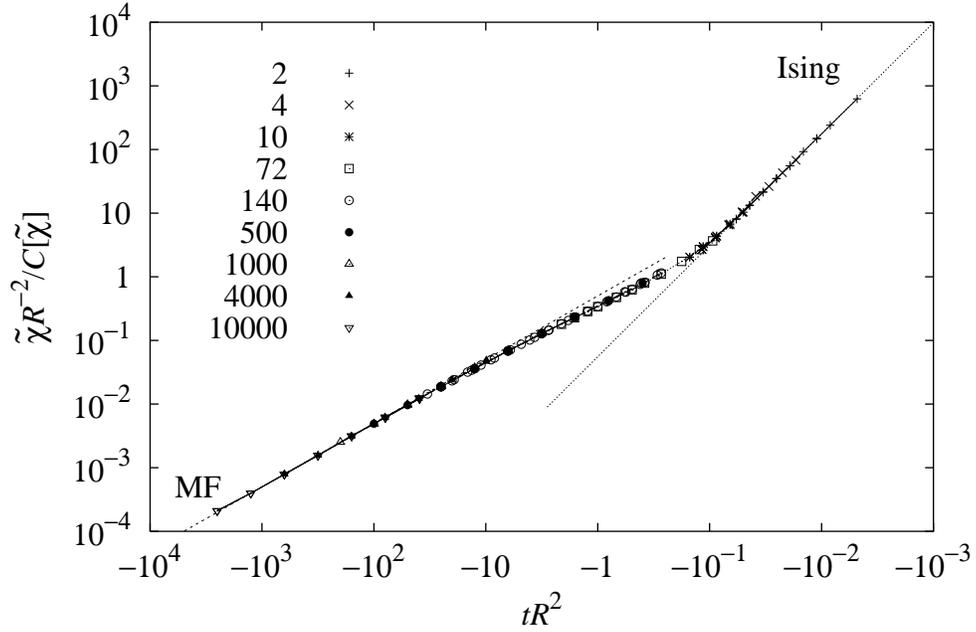} \\
\leavevmode (b)
\end{center}
\caption[]{Thermal crossover for the connected susceptibility~$\tilde{\chi}$
  for various ranges and system sizes. A finite-range correction
  factor~$C[\tilde{\chi}]$ has been divided out. Figure~(b) has also been
  corrected for saturation effects for $R_m^2 \geq 72$ and data points in the
  finite-size regime have been omitted.  For a discussion see the text.}
\label{fig:con-h}
\end{figure}

In Fig.~\ref{fig:con-h}(a) we show the connected susceptibility, appropriately
scaled with~$R$ and divided by the correction factor~$C[\tilde{\chi}]$, as a
function of the crossover variable.  Just as for the magnetization density,
deviations from the crossover curve are present even after the finite-range
corrections have been applied. These effects are either caused by finite-size
effects (close to~$T_{\rm c}$) or by systems that leave the critical region. In
the latter case, saturation effects start to come into play.  The finite-size
effects are clearly recognizable in the rightmost part of the graph, where the
curves start to follow horizontal lines. Once the temperature has been
sufficiently decreased, the graphs start following an asymptote with
slope~$-7/4$, on which the data for various ranges quite accurately collapse.
The amplitude of this asymptote is simply related to the exactly known
amplitude $A^{-}=0.025537\ldots$~\cite{barouch} of the {\em reduced\/}
susceptibility~$\chi_0$.  This reduced susceptibility is defined as $\chi_0
\equiv k_{\rm B}T\tilde{\chi} / \mu^2$, where $\mu$ denotes the magnetic moment
of a spin. This magnetic moment has also in our calculations implicitly been
divided out. However, we should keep in mind that we have expressed all
temperatures in terms of the mean-field critical temperature, i.e.\ $T_{\rm
  c}=1/(zK_{\rm c})$, where $z$ is the coordination number. For the
nearest-neighbor model this yields an additional factor~4 and we thus expect a
critical amplitude $2\arcsinh(1)A^{-}$. In addition we have to take into
account the finite-range correction factor which has been divided out. The
question whether this factor is applicable for $R=1$ has already been discussed
in Sec.~\ref{sec:mag-therm} [below Eq.~(\ref{eq:ising-asymptote})]. Here, the
difference between the deviation from the leading scaling behavior as predicted
by~$C[\tilde{\chi}]$ and the numerical result is approximately~3\%, whereas the
smallest differences that can be discerned on the logarithmic scale of
Fig.~\ref{fig:con-h}(a) are of the order of~5\%.  The asymptote with the
above-mentioned amplitude divided by~$C[\tilde{\chi}]$ indeed lies tangential
to the crossover curve, confirming our data. As the temperature is further
decreased, the data for systems with small interaction ranges start to follow a
line with a slope between that of the Ising and the mean-field asymptotes. This
effect is caused by the fact that these systems have left the critical region.
For sufficiently large interaction ranges, however, the curves coincide and
have a slope that gradually decreases (in the absolute sense). Although the
crossover curve at first varies more rapidly than for the magnetization
density, it subsequently only slowly approaches the classical regime and the
overall size of the crossover region is again between two and three decades.
Remarkably, the slope of the crossover curve passes even {\em through\/} the
mean-field value~$-1$ before settling at this value for sufficiently low
temperatures. In other words, the derivative of the connected susceptibility
appears to change nonmonotonically from its asymptotic Ising value~$-7/4$ to
its classical value~$-1$.  Several explanations may be considered for this
behavior. Either it is an intrinsic effect of the crossover function or it
might be explained from the fact that $\tilde{\chi}$ is the difference between
$\langle m^2 \rangle$ and~$\langle |m| \rangle^2$, which each separately are
described by a monotonically varying curve.

The saturation effects can---just as for the magnetization density---for large
ranges be described with mean-field theory. In a mean-field model the magnetic
susceptibility is given by
\begin{equation}
   \chi = \frac{1 - m^2}{t+m^2} \;.
\label{eq:chi-mf}
\end{equation}
Using Eq.~(\ref{eq:magseries}) we find for $T < T_{\rm c}$
\begin{equation}
   \chi = \frac{1}{-2t} - \frac{9}{10} + \frac{18}{175}(-t) +
          \frac{18}{175}(-t)^2  + \frac{6714}{67375}(-t)^3 +
          {\cal O}((-t)^4) \;,
\label{eq:chi-sat}
\end{equation}
which exhibits the classical value for the susceptibility exponent,
$\gamma_{\rm MF}=1$, and the critical amplitude~$\frac{1}{2}$.
Figure~\ref{fig:con-h}(a) shows the asymptote with this amplitude and one can
observe that the crossover curve approaches this asymptote from {\em below\/}
around $tR^2=-1$. Also the mean-field curves~(\ref{eq:chi-sat}) are shown for
$R_m^2=140$, $500$, $1000$, $4000$, and~$10000$ and they accurately describe
the numerical data. Thus, we have used the ratio between the series
expansion~(\ref{eq:chi-sat}) and the asymptotic behavior~$1/(-2t)$ to remove
the saturation effects in Fig.~\ref{fig:con-h}(a). The resulting graph is shown
in Fig.~\ref{fig:con-h}(b), in which also the data points in the finite-size
regime have been omitted in order to obtain a clear crossover curve. The
nonmonotonical variation of the slope of this curve is clearly visible.

\begin{figure}
\begin{center}
\leavevmode
\epsfbox{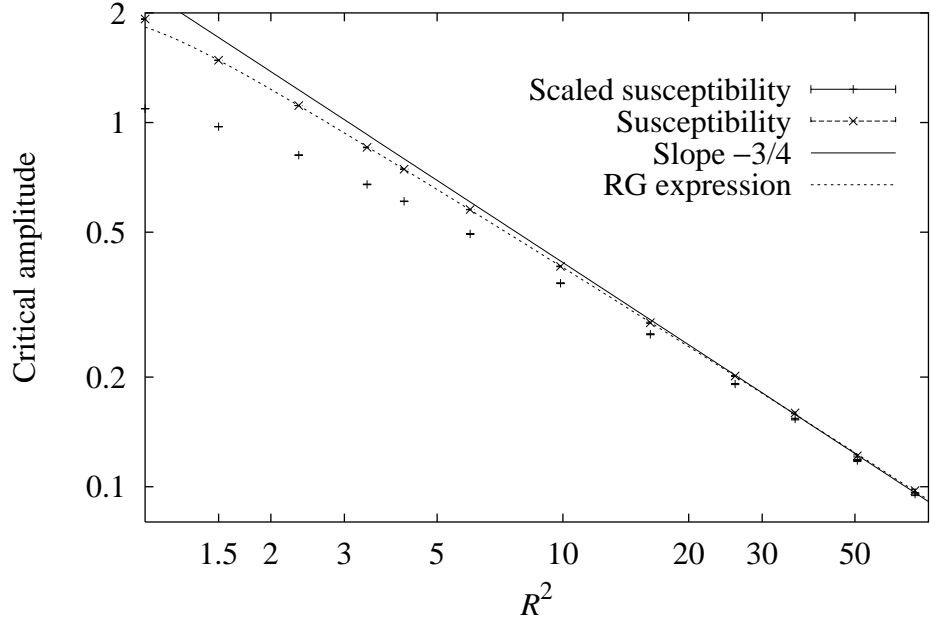}
\end{center}
\caption[]{Critical amplitude for the susceptibility $\chi' = L^d \langle m^2
  \rangle/k_{\rm B}T$ as extracted from the thermodynamic limit of
  $L^{-7/4}\chi'_L(K_{\rm c})$. The dashed curve indicates the renormalization
  prediction fitted to the numerical data.  Also the critical amplitude of the
  scaled susceptibility $k_{\rm B}T_{\rm c}\chi'$ is shown, which for small
  ranges deviates considerably stronger from the asymptotic behavior.}
\label{fig:chi-amp}
\end{figure}

\begin{figure}
\begin{center}
\leavevmode
\epsfbox{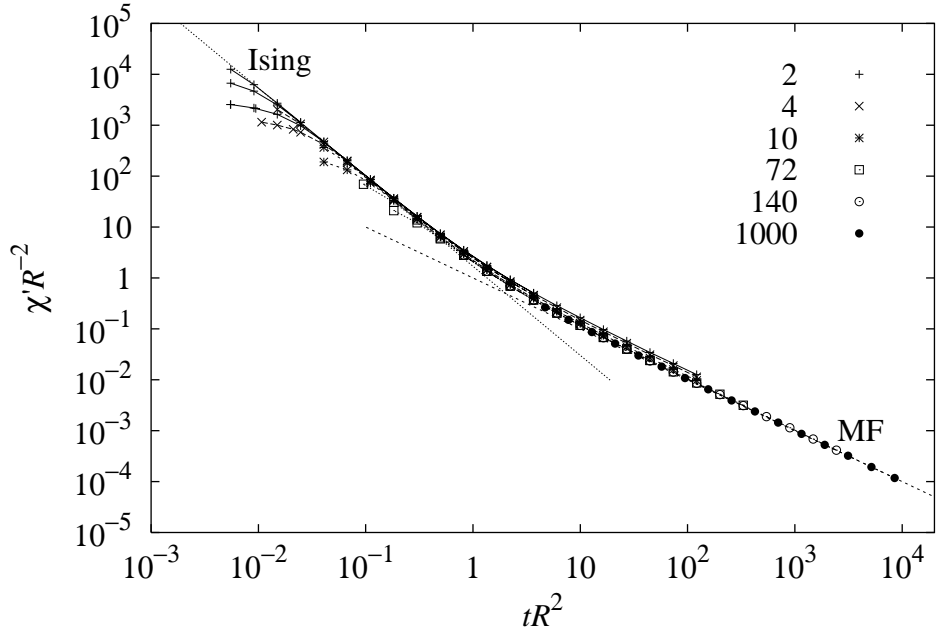}
\end{center}
\caption[]{Thermal crossover for the susceptibility~$\chi'$ in the symmetric
  phase, for various ranges and system sizes.  No finite-range corrections have
  been applied. For a discussion see the text.}
\label{fig:htchi}
\end{figure}

In the disordered (symmetric) phase, we encounter a different situation. The
susceptibility is now given by $\chi' \equiv L^d \langle m^2 \rangle / k_{\rm
  B}T$. This is identical to the expression we have used for the finite-size
crossover scaling, except that the temperature-dependent factor has been
omitted in Sec.~\ref{sec:chi-fss}. Figure~\ref{fig:chi-amp} shows the critical
finite-size amplitudes of both $\chi'$ and $\chi=L^d \langle m^2 \rangle$ as a
function of the interaction range.  We have fitted an expression of the
form~(\ref{eq:chi-scale}) to the data for $R_m^2 \geq 2$.  This expression
describes the data well, except for the data point at~$R_m^2=1$, where the
deviation is approximately~10\%.  Just as for the connected susceptibility, the
finite-range corrections to the critical amplitude of~$\chi'$ are much smaller
than for~$\chi$.  In fact, they are so small that they can be completely
omitted in the thermal crossover scaling, as illustrated in
Fig.~\ref{fig:htchi}. This graph shows $\chi'/R^2$ as a function of the
crossover variable~$tR^2$ for various interaction ranges and system sizes.
Outside of the finite-size regime, the data follow the Ising asymptote with
slope~$-7/4$.  The exactly known amplitude $2\arcsinh(1)A^{+}$, where
$A^{+}=0.96258\ldots$~\cite{barouch}, of this asymptote is accurately
reproduced by the numerical data. For larger temperatures, the curves gradually
approach an asymptote with the mean-field slope~$-1$.  However, some care has
to be exercised when interpreting this behavior.  Above~$T_{\rm c}$, no
saturation of the order parameter occurs and the system smoothly passes over to
regular (noncritical) behavior. In this high-temperature region the
susceptibility decreases proportional to~$1/T$.  For small interaction ranges
it is this behavior that one observes in the graph. Only for larger interaction
ranges one actually observes classical {\em critical\/} behavior. The latter
systems indeed reproduce the mean-field critical amplitude, which is equal
to~$1$ [as follows from Eq.~(\ref{eq:chi-mf}) with $m=0$]. Note that, due to
the absence of saturation effects, interaction ranges up to $R_m^2=1000$ are
amply sufficient to observe the full crossover region.

\section{Effective exponents}
\label{sec:effexp}
In several papers (see, e.g., Refs.~\cite{fisher-eff,anisimov95}) the slopes of
the crossover functions are described by so-called effective exponents. These
exponents can be defined as $\beta_{\rm eff} \equiv d \ln \langle m \rangle / d
\ln |t| = t\, d \ln \langle m \rangle / d t$ and $\gamma_{\rm eff} \equiv - d
\ln \chi / d \ln |t| = - t\, d \ln \chi / d t$. In fact, this concept is
familiar from the analysis of experimental data since a long
time~\cite{kouvel}, but only a limited amount of theoretical work has addressed
these issues.  Of course, these exponents change from their Ising values to the
classical values in the crossover region.  However, the precise variation in
the crossover region is unclarified and partially subject to debate. Although
these exponents can be read off from the form of the crossover curves presented
in the previous section, we consider it worthwhile to present separate graphs
displaying $\beta_{\rm eff}$, $\gamma_{\rm eff}^{-}$, and~$\gamma_{\rm
eff}^{+}$, where the superscripts denote the cases $t<0$ and $t>0$,
respectively. The additional advantage of these exponents is that they follow
from data with the same range and hence are not affected by any range-dependent
correction factors.

\begin{figure}
\begin{center}
\leavevmode
\epsfbox{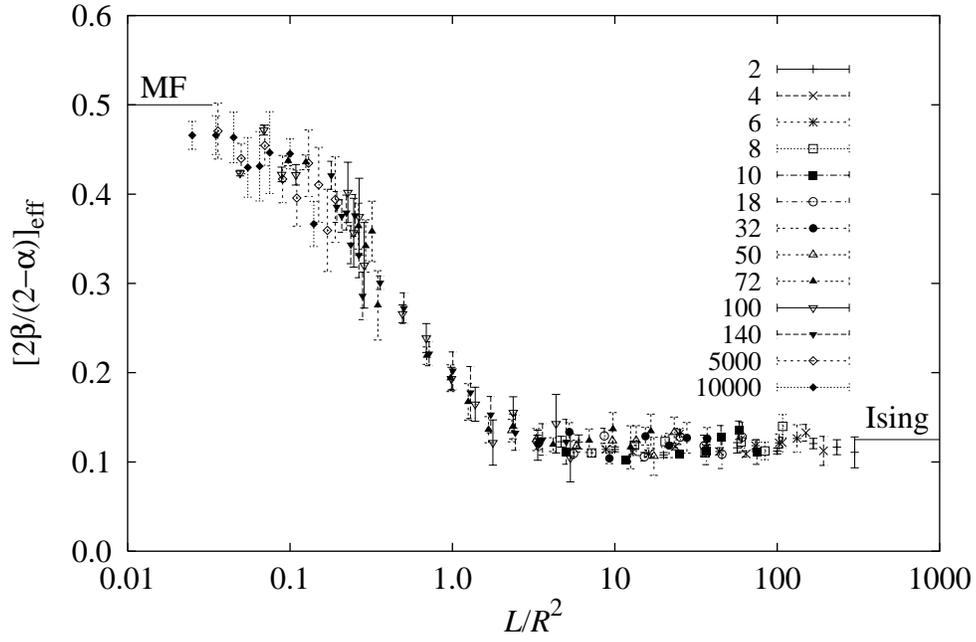}
\end{center}
\caption[]{The effective exponent $[2\beta/(2-\alpha)]_{\rm eff}$ as obtained
  from the finite-size crossover curve for $\langle |m| \rangle \sqrt{L}$.}
\label{fig:betanu-eff}
\end{figure}

As the graph in Fig.~\ref{fig:m-cross}(b) is particularly smooth, it is
tempting to consider its derivative as well. As derived in Ref.~\cite{medran},
$\langle |m| \rangle \sqrt{L} \propto (L/R^2)^{y_{\rm h}-3/2}$. This relation
also holds in the mean-field regime, provided that one replaces the magnetic
exponent~$y_{\rm h}$ by its starred counterpart $y_{\rm h}^* \equiv 3d/4 =
3/2$. The asterisk indicates that the exponent is modified due to the dangerous
irrelevant variable mechanism, as explained in, e.g., Ref.~\cite{univers}.
Thus, while we can rewrite the above-mentioned relation in the Ising regime in
terms of conventional critical exponents as $\langle |m| \rangle \sqrt{L}
\propto (L/R^2)^{-\beta/\nu + 1/2}$, this is not possible in the mean-field
regime, since $\nu_{\rm MF}$ is not affected by the dangerous irrelevant
variable mechanism. As an alternative we employ the specific heat
exponent~$\alpha$; $\langle |m| \rangle \sqrt{L} \propto
(L/R^2)^{-2\beta/(2-\alpha) + 1/2}$. The fact that the latter relation holds in
the mean-field regime while the former does not is a direct manifestation of
the violation of hyperscaling. Thus, we define $[2\beta/(2-\alpha)]_{\rm eff}
\equiv \frac{1}{2} -d \ln (\langle |m| \rangle \sqrt{L}) / d \ln (L/R^2)$. This
quantity is shown as a function of $L/R^2$ in Fig.~\ref{fig:betanu-eff}.
Although the error bars are considerable, the crossover from the Ising
value~$1/8$ (for large values of $L/R^2$) to the classical value~$1/2$ (for
small values of $L/R^2$) is clearly visible.

\begin{figure}
\begin{center}
\leavevmode
\epsfbox{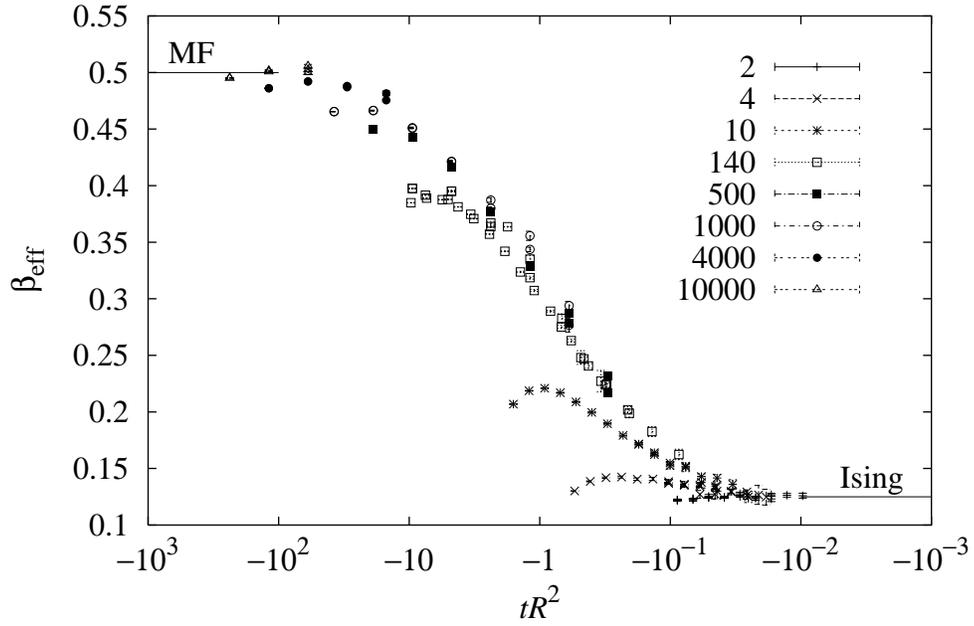}
\end{center}
\caption[]{The effective magnetization exponent $\beta_{\rm eff}$ describing
  the logarithmic derivative of the crossover function for the magnetization
  density.}
\label{fig:beta-eff}
\end{figure}

\begin{figure}
\begin{center}
\leavevmode
\epsfbox{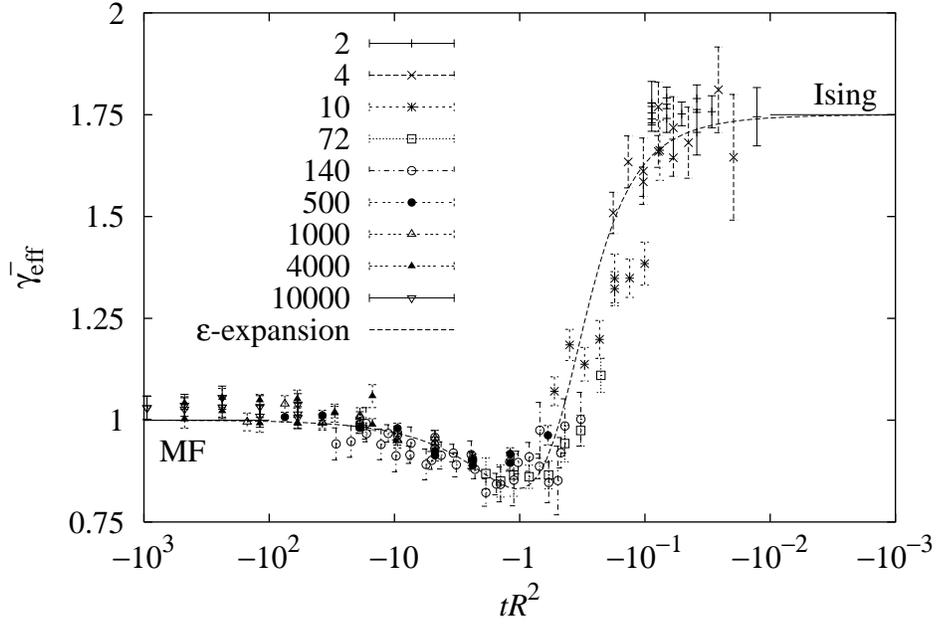}
\end{center}
\caption[]{The effective susceptibility exponent $\gamma_{\rm eff}^{-}$
  describing the logarithmic derivative of the crossover function for the
  connected susceptibility. The results on the left-hand side lie somewhat
  above the mean-field exponent due to saturation effects.}
\label{fig:gamma-eff-low}
\end{figure}

\begin{figure}
\begin{center}
\leavevmode
\epsfbox{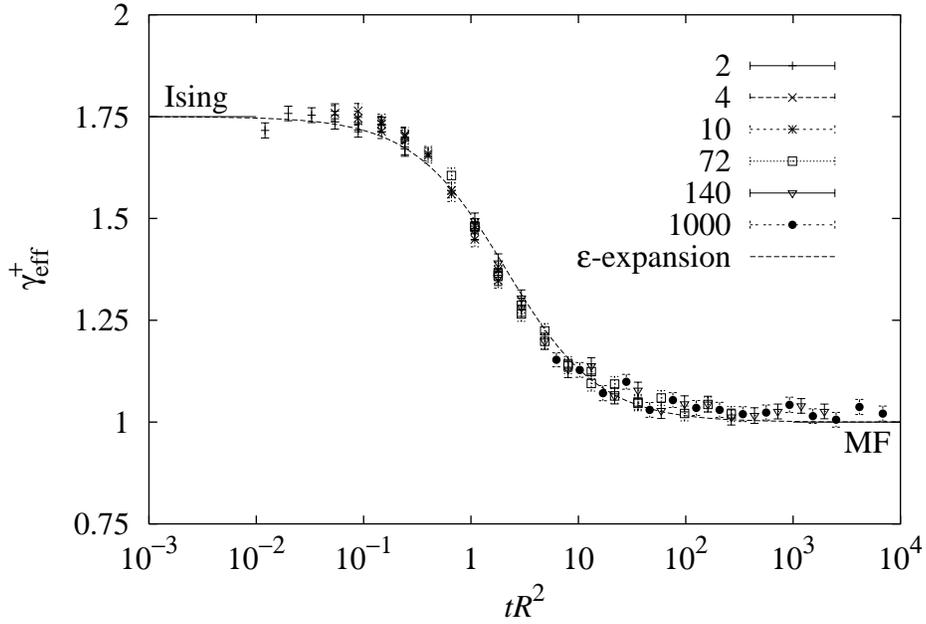}
\end{center}
\caption[]{The effective susceptibility exponent $\gamma_{\rm eff}^{+}$
  describing the logarithmic derivative of the crossover function for the
  susceptibility above~$T_{\rm c}$.}
\label{fig:gamma-eff-high}
\end{figure}

Turning to thermal crossover, we display in Figs.\ \ref{fig:beta-eff},
\ref{fig:gamma-eff-low}, and~\ref{fig:gamma-eff-high} the exponents $\beta_{\rm
eff}$, $\gamma_{\rm eff}^{-}$, and~$\gamma_{\rm eff}^{+}$ as defined above.
The effective magnetization exponent~$\beta_{\rm eff}$ increases monotonically
from its Ising value~$1/8$ to the classical value~$1/2$. In particular do the
data for different interaction ranges roughly fall onto the same curve, which
supports the hypothesis that the crossover curve is universal.  However, one
observes that for systems with relatively small interaction ranges the
effective exponent does {\em not\/} follow this curve.  This effect, caused by
saturation of the order parameter, can clearly lead to misleading results in
experiments!  In Fig.~\ref{fig:gamma-eff-low} the nonmonotonical variation of
$\gamma_{\rm eff}^{-}$ between $7/4$ and~$1$ is clearly visible. This may be
considered as a manifestation of what Fisher~\cite{fisher-eff} calls an
``underswing''. The occurrence of such a nonmonotonical crossover has been
predicted by various renormalization calculations for the crossover from Ising
to $XY$ and Heisenberg critical behavior {\em above\/}~$T_{\rm c}$, see, e.g.,
Refs.~\cite{nelson,amit,seglar} and references therein. Furthermore, an
exponent $\gamma_{\rm eff}=0.88(3)$ has been measured in the symmetric phase in
micellar solutions~\cite{corti}. Fisher~\cite{fisher-eff} has suggested that an
effective susceptibility exponent that takes a value $\gamma_{\rm eff}<1$ in
the crossover region might be a general feature of crossover from 3D Ising to
classical critical behavior and noted that concrete calculations yielding such
an effective exponent would be valuable. In Ref.~\cite{fisher-eff}, a
first-order $\epsilon$-expansion is quoted for the exponent crossover function,
\begin{equation}
\gamma_{\rm eff} = 1 + (\gamma_{\rm I}-\gamma_{\rm MF}) E[\ln(|t/G|)] \;,
\label{eq:gamma-eff}
\end{equation}
where $G$ is the crossover temperature or Ginzburg number and
\begin{equation}
E(\ln y) = 1/(1+y^{\epsilon/2}) \;.
\label{eq:exp-crossover}
\end{equation}
In our case, $t/G$ is directly proportional to the crossover variable~$tR^2$.
To describe the experimental results from Ref.~\cite{corti}, Fisher used the
following extension of~Eq.~(\ref{eq:exp-crossover}),
\begin{equation}
E(\ln y) = (1+py^{\epsilon/2})/[1+(p+1)y^{\epsilon/2}+qy^{\epsilon}] \;.
\label{eq:approx-exp}
\end{equation}
Even though one may not expect such an expansion to converge for $d=2$, we have
drawn expression~(\ref{eq:gamma-eff}) in Fig.~\ref{fig:gamma-eff-low}, where we
have taken the function $E(\ln y)$ from Eq.~(\ref{eq:approx-exp}), set
$\epsilon=2$ and adjusted $p$ and~$q$ such that the curve constituted a
reasonable description of the data. Clearly, no conclusions should be drawn
from this curve, especially because Eq.~(\ref{eq:approx-exp}) has been proposed
for the symmetric phase. In addition, for $d=2$ the exponent~$\epsilon/2$ is a
very poor approximation for the exponent $\theta \equiv -y_{\rm i}/y_{\rm
t}=2$, which is actually expected to appear in the function~$E(\ln y)$.  As
follows from Fig.~\ref{fig:gamma-eff-high}, the behavior above~$T_{\rm c}$ is
completely different. Here we have used expression~(\ref{eq:gamma-eff}) with
Eq.~(\ref{eq:exp-crossover}) to describe the data. Except for a shift along the
horizontal axis (a proportionality constant in the Ginzburg number), no
adjustable parameter is present and it is surprising how well the data agree
with the theoretical prediction.

Sometimes experiments have yielded effective exponents in disagreement with the
known~\cite{fisher-review} universality classes, but still satisfying the
scaling relations, such as $\gamma_{\rm eff}+2\beta_{\rm eff}=2-\alpha_{\rm
eff}$. Here $\alpha_{\rm eff}$ denotes the effective exponent of the specific
heat, which in our case is expected to be always (close to) zero, as both the
classical and the 2D Ising value of $\alpha$ are equal to zero. This is also
confirmed by the close resemblance between Figs.\ \ref{fig:betanu-eff}
and~\ref{fig:beta-eff}. Thus, it is interesting to note that this scaling
relation is strongly violated in the present case: from Figs.\
\ref{fig:beta-eff} and~\ref{fig:gamma-eff-low} we can estimate that
$\gamma_{\rm eff}+2\beta_{\rm eff}$ reaches a minimum of approximately~$1.4$ at
$tR^2 \approx -1$.

\section{Conclusions}
\label{sec:conclusions}
In this paper we have presented numerical results for scaling functions
describing the crossover from Ising-like to classical critical behavior in
two-dimensional systems. While the general concepts describing this crossover
have been developed many years ago, only a limited amount of progress has been
made for a long time. In the present paper it is demonstrated, for the first
time, that one can obtain accurate quantitative information on crossover
scaling from computer simulations. The full crossover region was covered both
for finite-size crossover and thermal crossover above and below~$T_{\rm c}$. A
data collapse has been obtained for all system sizes and interaction ranges,
which supports the hypothesis that these crossover functions are universal.
Deviations from this curve are present but can be understood from finite-size
and saturation effects. The results are in agreement with the previously
derived renormalization scenario for these systems.

Working in two dimensions offers the advantage that the exponents and the
critical amplitudes are known exactly. More importantly, critical fluctuations
are very large in two dimensions, which leads to critical behavior that
strongly differs from classical behavior and hence to a clearly visible
crossover between the two universality classes. We have shown that the
magnetization density is described by a smooth crossover curve.  The effective
exponent, defined as the logarithmic derivative of this curve, increases
monotonically from the Ising value to the classical value in two or three
decades in the reduced temperature. On the other hand, the effective exponent
for the susceptibility has a logarithmic derivative which varies monotonically
above the Curie temperature and nonmonotonically below it. The occurrence of
nonmonotonic behavior in the symmetric phase has been inferred from
renormalization-group calculations in three dimensions and found long-standing
interest.  An extension of the present study to $d=3$ is therefore highly
desirable and has been planned for the near future.

\acknowledgments
E.~L. and H.~B. acknowledge the kind hospitality of the condensed matter theory
group of the Johannes Gutenberg-Universit\"at Mainz, where part of this work
has been completed. We are grateful to the H\"ochstleistungsrechenzentrum
(HLRZ) J\"ulich for access to a Cray-T3E, on which part of the computations
have been carried out.

\appendix
\section*{Exact calculation of some finite-size effects in a mean-field system}
In Ref.~\cite{ijmpc} the universal amplitude ratio $Q$ has been calculated for
a system in which all spins interact equally strongly, including the leading
finite-size correction. It was shown that the relevant integrals can be
expressed in terms of a quantity $I_k$, which we here generalize to odd powers
of the magnetization density,
\begin{equation}
I_k \equiv
      \int_{-\infty}^{\infty} dm\; |m|^k \exp\left(-\frac{1}{12}Nm^4\right)
    = \left(\frac{12}{N}\right)^{\frac{k+1}{4}}\frac{1}{2}
      \Gamma\left(\frac{k+1}{4}\right) \;,
\end{equation}
where $N$ denotes the number of spins. In a similar fashion we can also
calculate other finite-size effects to leading order in~$N$.
Expanding the terms in the partition function and replacing the sum over all
possible states by an integral, we find the following expression for the
absolute magnetization density
\begin{equation}
\langle |m| \rangle =
 \frac{1}{Z}\int_{-\infty}^{\infty} dm\; |m| \exp\left(-\frac{1}{12}Nm^4\right)
  \left[ 1 - \frac{1}{30}Nm^6 + \frac{1}{2}m^2 + {\cal O}(Nm^8,m^4)\right] \;,
\end{equation}
in which $Z$ denotes the partition function (except for a prefactor which has
been divided out, cf.\ Eq.~(31) in Ref.~\cite{ijmpc}). Elementary algebra leads
then to
\begin{equation}
\langle |m| \rangle =
    12^{1/4}\frac{\Gamma (\frac{1}{2})}{\Gamma (\frac{1}{4})}\frac{1}{N^{1/4}}
    + {\cal O}\left(\frac{1}{N^{3/4}}\right) \;,
\end{equation}
which for $d=2$ yields expression~(\ref{eq:mag-mf}). Along the same lines  one
finds
\begin{equation}
\langle m^2 \rangle =
    \sqrt{12}\frac{\Gamma (\frac{3}{4})}{\Gamma (\frac{1}{4})}\frac{1}{N^{1/2}}
    + {\cal O}\left(\frac{1}{N}\right) \;,
\end{equation}
i.e.\ the susceptibility diverges as~$\sqrt{N}$, and
\begin{equation}
\langle m^4 \rangle =
       12 \frac{\Gamma (\frac{5}{4})}{\Gamma (\frac{1}{4})}\frac{1}{N}
       + {\cal O}\left(\frac{1}{N^{3/2}}\right)
    =  \frac{3}{N} + {\cal O}\left(\frac{1}{N^{3/2}}\right) \;.
\end{equation}

\newpage
% figures follow here

% tables follow here
\newpage
%\pagebreak[2]

\begin{table}
\renewcommand{\arraystretch}{1.3}
\begin{tabular}{r|r|r|r|r}
$R_m^2$  & $z$ & $R^2$ & $K_{\rm c}$ & $K_{\rm c}^{\rm MF}$ \\ \hline
  500  &  1580 & $\frac{99449}{395}    \approx 251.770$
               & $6.379 (2) \times 10^{-4}$   & $6.3291139 \times 10^{-4}$ \\
 1000  &  3148 & $\frac{394530}{787}   \approx 501.309$
               & $3.1904 (6) \times 10^{-4}$  & $3.1766201 \times 10^{-4}$ \\
 4000  & 12580 & $\frac{1259568}{629}  \approx 2002.49$
               & $7.9594 (5) \times 10^{-5}$  & $7.9491256 \times 10^{-5}$ \\
 5000  & 15704 & $\frac{9813759}{3926} \approx 2499.68$
               & $6.3746 (3) \times 10^{-5}$  & $6.3678044 \times 10^{-5}$ \\
10000  & 31416 & $\frac{6545445}{1309} \approx 5000.34$
               & $3.18491 (9) \times 10^{-5}$ & $3.1830914 \times 10^{-5}$ \\
\end{tabular}
\caption{Some properties of the additional ranges used to span the full thermal
  crossover region. $R_m^2=5000$ has been included for completeness; it has
  only been used for the finite-size crossover scaling. The first three columns
  list the squared range of interaction~$R_m^2$, the corresponding number of
  neighbors~$z$, and the squared effective range of interaction~$R^2$.
  Furthermore the critical coupling~$K_{\rm c}$ as calculated from
  Eq.~(\protect\ref{eq:tc-scale}) and the mean-field approximation for the
  critical coupling $K_{\rm c}^{\rm MF}=1/z$ are shown.}
\label{tab:largerange}
\end{table}

\end{document}